\documentclass[prx, aps, 10pt, twocolumn, amsmath, amssymb, superscriptaddress, showpacs, nofootinbib]{revtex4-2}

\usepackage{amsmath}
\usepackage{graphicx}
\usepackage{verbatim}
\usepackage{color}
\usepackage[dvipsnames,svgnames,table]{xcolor}
\usepackage{subfigure}
\usepackage{multirow}
\usepackage{float}
\usepackage{braket}
\usepackage{siunitx}
\usepackage{bbm}
\usepackage{bibunits}
\usepackage{bbold}
\usepackage{tikz}
\usepackage[inkscapeformat=pdf, inkscapearea=page,
			inkscapelatex=false,
]{svg}
\usepackage[hidelinks]{hyperref}
\usepackage[capitalize]{cleveref}
\usepackage[all]{hypcap}
\usepackage{microtype}

\newcommand{\e}[1]{\ensuremath{\mathrm e^{#1}}}
\newcommand{\diff}{\mathrm{d}}

\newcommand{\affiliationeth}{\affiliation{Department of Physics, ETH Zurich, CH-8093 Zurich, Switzerland}}
\newcommand{\affiliationqc}{\affiliation{Quantum Center, ETH Zurich, CH-8093 Zurich, Switzerland}}
\newcommand{\affiliationpsi}{\affiliation{ETH Zurich - PSI Quantum Computing Hub, Paul Scherrer Institute, CH-5232 Villigen, Switzerland}}
\newcommand{\figpanel}[1]{\textbf{#1}}
\newcommand{\figpanelref}[1]{\figpanel{#1}}
\newcommand{\figtitle}[1]{#1}

\newcommand{\qbh}{\mathrm{Q_{high}}}
\newcommand{\qbl}{\mathrm{Q_{low}}}
\newcommand{\fluxh}{\phi_{\mathrm{high}}}
\newcommand{\fluxl}{\phi_{\mathrm{low}}}
\newcommand{\cphase}{\theta}
\newcommand{\cphasei}{\theta_i}
\newcommand{\kphase}{\theta_{\mathrm{idle}}}
\newcommand{\translength}{t_{\mathrm{idle}}}
\newcommand{\pulselength}{t_{\mathrm{int}}}
\newcommand{\bufferlength}{t_{\mathrm{sep}}}
\newcommand{\deltat}{\Delta(t)}
\newcommand{\deltamin}{\Delta_{\mathrm{min}}}
\newcommand{\deltamax}{\Delta_{\mathrm{max}}}
\newcommand{\umix}{U_\mathrm{mix}}
\newcommand{\leakage}{L}
\newcommand{\jhalfo}{g_1}
\newcommand{\jhalft}{g_2}
\newcommand{\subsqbh}{\mathrm{high}}
\newcommand{\subsqbl}{\mathrm{low}}
\newcommand{\cz}{\mathrm{C}Z}

\begin{document}

\title{Realizing a Continuous Set of Two-Qubit Gates Parameterized by an Idle Time}
\author{Colin~Scarato}\thanks{Contact author: \href{mailto:colin.scarato@phys.ethz.ch}{colin.scarato@phys.ethz.ch}}\affiliationeth\affiliationqc
\author{Kilian~Hanke}\affiliationeth
\author{Ants~Remm}\thanks{Present address: Atlantic Quantum, Cambridge, MA}\affiliationeth
\author{Stefania~Laz\u{a}r}\thanks{Present address: Zurich Instruments AG, Zurich, Switzerland}\affiliationeth
\author{Nathan~Lacroix}\affiliationeth\affiliationqc
\author{Dante~Colao~Zanuz}\affiliationeth\affiliationqc
\author{Alexander~Flasby}\affiliationeth\affiliationqc\affiliationpsi
\author{Andreas~Wallraff}\affiliationeth\affiliationqc\affiliationpsi
\author{Christoph~Hellings}\affiliationeth\affiliationqc
\date{March 14, 2025}

\begin{abstract}

Continuous gate sets are a key ingredient for near-term quantum algorithms.
Here, we demonstrate a hardware-efficient,
continuous set of controlled arbitrary-phase ($\cz_{\cphase}$) gates
acting on flux-tunable transmon qubits.
This implementation is robust to control pulse distortions on time scales longer than the duration of the gate,
making it suitable for deep quantum circuits.
Our calibration procedure makes it possible to parameterize
the continuous gate set with a single control parameter,
the idle time between the two rectangular halves of the net-zero control pulse.
For calibration and characterization, we develop a leakage measurement based on
coherent amplification, and a new cycle design for cross-entropy benchmarking.
We demonstrate gate errors of $\SI{0.7}{\%}$ and leakage of $4\times 10^{-4}$ across the entire gate set.
This native gate set has the potential to
reduce the depth and improve the performance
of near-term quantum algorithms
compared to decompositions into $\cz_{\pi}$ gates and single-qubit gates.
Moreover, we expect the calibration and benchmarking methods to find further possible applications.

\end{abstract}

\maketitle

\section{Introduction}

Variational quantum algorithms on superconducting qubits
are under investigation for applications as diverse as
quantum chemistry \cite{Bauer2020, Kandala2017},
simulations of quantum many-body systems \cite{Barends2015, Morvan2022b},
quantum phase transitions \cite{Herrmann2022},
metrology \cite{Marciniak2022},
machine learning \cite{Schuld2018c, Havlicek2019},
solving optimization problems \cite{Otterbach2017, Harrigan2021},
and foundational aspects of quantum information theory \cite{Jafferis2022, Morvan2024}.
They typically include continuously parameterized unitaries \cite{Farhi2014, Peruzzo2014, Farhi2018}, which may lead to prohibitive circuit depths if compiled into hardware operations from a minimal universal gate set \cite{Vatan2004}.
Hardware-efficient implementations of such parameterized unitaries, in the form of continuous gate sets, allow to reduce the depth of near-term, noisy quantum circuits and improve their performance \cite{Lacroix2020, Cirstoiu2020, PerezSalinas2023, Preti2024}.

Continuous gate sets have been demonstrated for SWAP-like gates \cite{Ganzhorn2019, Foxen2020},
$XY$ gates \cite{Abrams2020},
three-qubit unitaries \cite{Roushan2016, Hill2021},
as well as for controlled arbitrary-phase ($\cz_{\cphase}$) gates,
which apply a conditional phase $\cphase$ to the $\ket{11}$ state \cite{Barends2015, Lacroix2020, Collodo2020, Foxen2020}.
While high $\cz_{\cphase}$ gate fidelities have been demonstrated with tunable qubits in combination with tunable couplers \cite{Foxen2020},
the latter come at the cost of additional wiring and on-chip elements for the flux control of the couplers as well as calibration overhead.
This increased complexity is avoided in architectures with fixed coupling between flux-tunable transmon qubits.
Controlled-phase gates \cite{DiCarlo2009} or controlled arbitrary-phase gates \cite{Barends2015,Lacroix2020}
can then be implemented
by using flux pulses to realize a direct resonant exchange between the
$\ket{11}$ state and the non-computational $\ket{20}$ state.
However, these implementations are susceptible to
control pulse distortions on time scales longer than the gate duration,
affecting subsequent gates in long gate sequences,
an effect termed ``nonatomicity'' \cite{Rol2019} or ``bleeding'' \cite{Foxen2020}.

To provide robustness against these distortions, controlled-phase gates
were implemented based on net-zero waveforms with vanishing zero-frequency component \cite{Rol2019, Negirneac2021}.
In this scheme, the flux pulses
are split into two halves with opposite polarities and an idle time between them.
During the idle time, the population temporarily brought to the $\ket{20}$ state acquires a phase,
which directly contributes to the final conditional phase $\cphase$.
However,
the calibration procedures in \cite{Rol2019, Negirneac2021} focus on a fixed conditional phase $\cphase=\pi$ and
do not suffice to calibrate a whole continuous gate set.

Here, we extend the net-zero $\cz_{\pi}$ gate from \cite{Negirneac2021} to a continuous set of $\cz_{\cphase}$ gates.
We introduce a novel calibration procedure,
which leads to a
simple linear dependence of the conditional phase $\cphase$
on a single control parameter, the idle time $\translength$.
This is in contrast to previous implementations of $\cz_{\cphase}$ gate sets,
where tuning through all phases $\cphase$ required adjusting multiple interdependent parameters
\cite{Barends2015,Foxen2020,Lacroix2020}.
We fine-tune the $\ket{11}\leftrightarrow\ket{20}$ resonance
beyond the reach of the existing method from \cite{Negirneac2021},
a requirement to attain low leakage
to the $\ket{20}$ state
across the whole continuous gate set.

Using cross-entropy benchmarking (XEB) \cite{Boixo2018, Arute2019, Helsen2022a},
we demonstrate that the gate performance is uniform over the whole range of phases $\cphase$.
Moreover, by applying pulses to both qubits simultaneously as in \cite{Herrmann2022, Krinner2022},
we realize the continuous gate set between qubits which are far-detuned by over $\SI{2}{\giga\hertz}$ when the gate is off,
suppressing residual $ZZ$ interactions
to below $\SI{5}{\kilo\hertz}$
despite a coupling strength of $\SI{11.38\pm0.01}{\mega\hertz}$ when the gate is on.
This on-off ratio is more than an order of magnitude higher than reported in implementations
where flux pulses are applied to only one qubit \cite{Lacroix2020, Negirneac2021}.

Two methods developed
in this work
are widely applicable beyond the context of
calibrating and characterizing continuous gate sets.
The first method is a leakage measurement based on coherent population amplification, which
can be beneficial for applications requiring the calibration of low-leakage gates
for quantum error correction \cite{Aliferis2007b, Fowler2013, Ghosh2015a, Suchara2015, Bultink2020, Varbanov2020, Battistel2021},
for example.
The second method is an XEB cycle design which,
as opposed to existing protocols \cite{Helsen2022a, Boixo2018, Sheldon2015, Erhard2019, Helsen2022},
allows to benchmark individual weakly entangling gates.

Below, we introduce the concept of the continuous set of controlled arbitrary-phase gates.
We then describe methods to fine-tune the resonance condition, necessary to reach low leakage uniformly for all phases.
Finally, we present the calibration procedure and benchmarking results for our implementation.

\section{Parameterization of the continuous gate set} \label{section:concept}

\begin{figure}
	\centering
	\includegraphics[width=\linewidth]{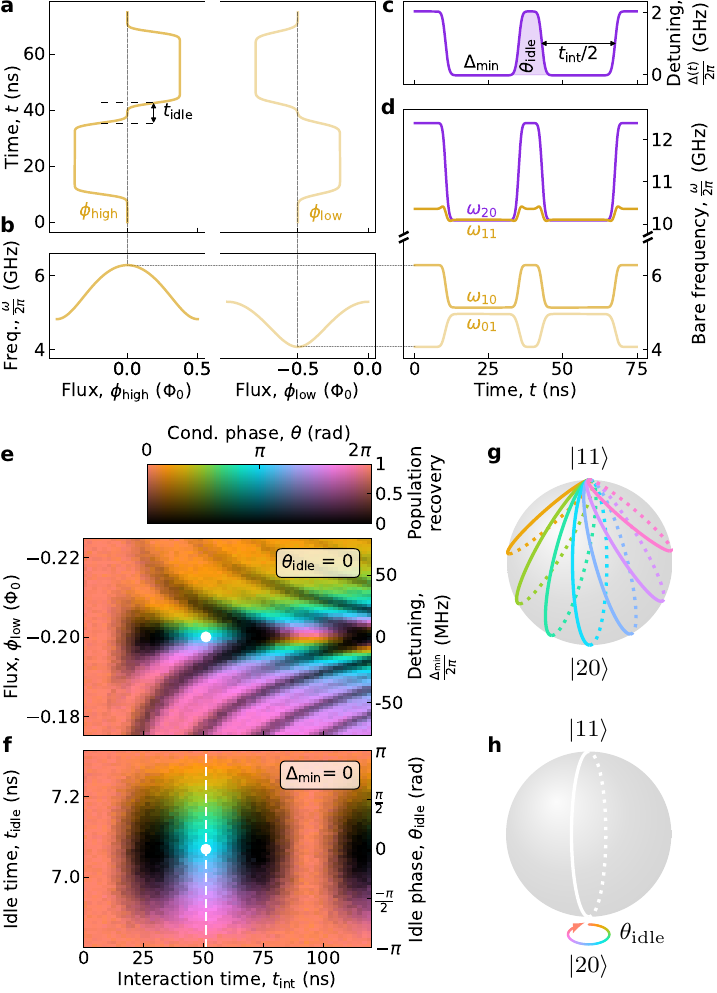}
	\caption{
		\figtitle{Concept of the gate set.}
		\figpanel{a}~Control pulses, as a function of time (vertical axis), in units of magnetic flux at the SQUID loop of each transmon.
		\figpanel{b}~Qubit transition frequencies $\omega_{10}$ and $\omega_{01}$ of the two transmons as a function of flux.
		\figpanel{c,d}~Frequencies of the relevant two-transmon states (\figpanelref{d}) and detuning $\deltat = \omega_{20}-\omega_{11}$ (\figpanelref{c}) over time due to the flux pulses in \figpanelref{a} and the conversion to frequencies in \figpanelref{b}. The shaded area in \figpanelref{c} indicates the phase $\kphase$ acquired by the $\ket{20}$ state during the idle time $\translength$.
		\figpanel{e,f}~Measured population (brightness) and conditional phase (hue) of the $\ket{11}$ state at the end of the gate,
		see measurement techniques in the main text.
		Shown are two cuts through the 3-dimensional parameter space ($\deltamin$, $\kphase$, $\pulselength$), at $\kphase=0$ (\figpanelref{e}) and $\deltamin=0$ (\figpanelref{f}).
		The white point indicates a $\cz_{\pi}$ gate, the dashed line the continuous gate set.
		\figpanel{g}~Bloch sphere illustrating state trajectories for several $\deltamin$ yielding $\cphase\in\{\pi/4, \pi/2, ..., 7\pi/4\}$, for $\kphase=0$ as in \figpanelref{e} (idealized illustration assuming pulses with zero rise time).
		\figpanel{h}~Illustration of the implementation of the continuous gate set by means of a variable idle phase as in \figpanelref{f}, see main text.
	}
	\label{fig:1}
\end{figure}

We work with two neighboring qubits in a six-transmon quantum device,
see \cref{supp:setup} for a summary of system parameters.
We write $\qbh$ to refer to the high-frequency qubit,
idling at $\SI{6.3}{\giga\hertz}$ in this particular experiment,
and $\qbl$ to refer to the low-frequency qubit,
idling at $\SI{4.1}{\giga\hertz}$.
The two qubits can be tuned in frequency with individual flux lines,
realized as coplanar waveguides shorted near the superconducting quantum interference device (SQUID) of each qubit.
Gaussian-filtered, piece-wise constant net-zero flux pulses generated by an arbitrary waveform generator (\cref{fig:1}a),
translate, via the flux-frequency characteristic of the transmons (\cref{fig:1}b),
into a time dependence of all the energy levels (\cref{fig:1}d).
We describe the measurement setup in \cref{supp:setup},
and details of the flux waveform generation in \cref{supp:waveform}.
To perform a two-qubit gate, we bring the $\ket{11}$ state into resonance with the $\ket{20}$ state (\cref{fig:1}d),
where the first number in the state label refers to the energy level of $\qbh$ and the second one corresponds to $\qbl$.
The time dependence of the detuning $\deltat = \omega_{20}-\omega_{11}$ (\cref{fig:1}c)
shows two plateaus at the minimal detuning $\deltamin\approx0$,
during which the resonant interaction is switched on,
separated by a time interval at maximum detuning,
where the qubits are back at their idle frequencies.

We express the Hamiltonian of the coupled system in a frame rotating at the qubit frequencies.
This amounts to canceling the single-qubit dynamic phases
accumulated by the frequency excursions shown in \cref{fig:1}d.
When performing the experiment, this cancellation is implemented as a redefinition of the qubit frames,
see \cref{supp:calib}.
In the two-dimensional subspace spanned by $\ket{11}$ and $\ket{20}$,
the Hamiltonian reduces to
\begin{align}
	\label{eq:Hmain}
	H(t)=
	\left(\begin{array}{c@{}}
		\begin{matrix}
			0&\jhalft/2\\
			\jhalft/2&\deltat\\
		\end{matrix}
	\end{array}\right)
\end{align}
where $\jhalft$ is the coupling rate in the two-excitation subspace, which is approximately independent of $\deltat$
as shown in \cref{supp:setup}.
During each half of the resonant interaction (the two plateaus in \cref{fig:1}c),
the $\ket{11}$ population is diabatically swapped to the $\ket{20}$ state,
and then swapped back to $\ket{11}$.
During the idle time $\translength$ while $\deltat\gg \jhalft$,
the $\ket{20}$ state acquires a phase $\kphase = \int_{t}^{}\deltat\diff t$
(shaded area in \cref{fig:1}c).

To visualize the parameter space,
we measure the population and phase of the $\ket{11}$ state after the two-qubit interaction
as a function of the pulse parameters.
Keeping the flux amplitude $\fluxh$ of $\qbh$ fixed,
we sweep the interaction time $\pulselength$,
together with the flux amplitude $\fluxl$, which sets the minimum detuning $\deltamin$ (\cref{fig:1}e),
or with the idle time $\translength$, which sets the idle phase $\kphase$ (\cref{fig:1}f).
We measure the population recovery (shown as brightness),
by applying the flux pulses to the qubits prepared in the $\ket{11}$ state
and performing averaged three-level readout \cite{Magnard2018},
see detailed method in \cref{supp:calib}.
The brightest points in \cref{fig:1}f correspond to quantum logic gates,
that is, unitaries retaining the state within the qubit subspace.
All darker areas correspond to unitaries with imperfect population recovery, that is, leakage to $\ket{20}$.
We measure the conditional phase (shown as hue)
with the Ramsey-type experiment described in \cref{supp:calib}.

The conditional phase can be intuitively understood as a
non-adiabatic generalization of the Berry phase (Aharonov-Anandan phase \cite{Aharonov1987}),
resulting from the geometrical trajectory on the $\ket{11}$--$\ket{20}$ Bloch sphere.
Geometrical phases have previously been explored
to implement quantum gates
in nuclear magnetic resonance qubits \cite{Wang2001}, spin qubits \cite{Solinas2003a}, trapped ions \cite{Duan2001},
and in superconducting qubits for single-qubit gates \cite{Blais2003},
controlled-phase gates in continuous-variable systems \cite{Song2017a},
and microwave-driven controlled-phase gates \cite{Xu2020g}.
See \cref{supp:geom_phase} for a geometrical description of the conditional phase in our gate scheme.

If we fix the idle phase $\kphase=0$,
as in \cref{fig:1}e,
we can define a continuous gate set along the curve of full population recovery
passing through the $\cz_{\pi}$ gate (white dot).
In this case,
the conditional phase $\cphase$ corresponds to the solid angle enclosed by the trajectory on the Bloch sphere,
as shown for a few choices of the detuning $\deltamin$ in \cref{fig:1}g
and discussed in detail in \cref{supp:geom_phase}.
However, since the interaction time $\pulselength$ needs to be adjusted
for any chosen detuning $\deltamin$ such that the trajectory ends in the $\ket{11}$ state,
a continuous gate set defined in this way is parameterized by two interdependent parameters
that need to be adjusted simultaneously.
If, instead of fixing $\kphase$, we fix the detuning $\deltamin=0$
as in \cref{fig:1}f,
the points with full population recovery form a pattern of straight lines,
at either constant interaction time $\pulselength$ (vertical lines) or constant idle phase $\kphase$ (horizontal lines).
Along the lines $\kphase=\pm\pi$,
the population recovery is maximized, but the conditional phase is zero,
which can be understood as an echo effect caused by the idle phase $\kphase$,
see Appendix A.
By contrast,
along the dashed white line of constant interaction time,
any conditional phase can be achieved,
making it possible to realize a continuous gate set by adjusting a single parameter.

As detailed in \cref{supp:geom_phase},
the conditional phase $\cphase$ of this continuous gate set can be expressed as
$\cphase = \pi-\kphase$.
The first term is a constant geometrical phase set by the trajectory on the Bloch sphere,
which is independent of $\kphase$,
as visible in \cref{fig:1}h.
The second term is a variable dynamic phase
set by the integral of the detuning,
$\kphase = \int_{t}^{}\deltat\diff t$.
This integral depends linearly on the idle time $\translength$,
provided that $\translength$ is long compared to the rise time of the pulses
such that $\deltat$ forms a plateau (\cref{fig:1}c).
This means that the conditional phase can be controlled linearly with the idle time $\translength$ while keeping all other parameters fixed.
In practice, we vary $\translength$ with sub-picosecond precision
by adjusting the amplitudes at the smooth pulse edges (\cref{supp:waveform}).

\section{Accurately establishing resonance between \texorpdfstring{$\ket{11}$}{11} and \texorpdfstring{$\ket{20}$}{20}} \label{section:resonance}

\begin{figure}
	\centering
	\includegraphics[width=\linewidth]{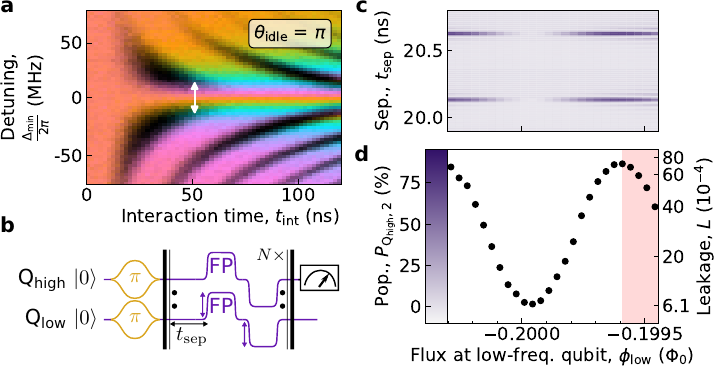}
	\caption{
		\figtitle{Characterization of the resonance condition.}
		\figpanel{a}~Population recovery and conditional phase (color scheme as in \cref{fig:1})
		as a function of the detuning $\deltamin$ (controlled by the pulse amplitude $\fluxl$) and interaction time $\pulselength$.
		For these measurements, $\kphase$ is set to $\pi$,
		resulting in maximum population sensitivity to $\deltamin$ (illustrated by the white arrow).
		\figpanel{b}~Pulse sequence for coherent amplification of leakage,
		by repeating $N$ times the flux pulses (FP) implementing the gate.
		Repetition is denoted by vertical bars with dots,
		and $\bufferlength$ is the time between centers of the falling and rising edges of successive gates.
		\figpanel{c}~Second excited state population, $P_{\qbh,2}$, of $\qbh$,
		color-coded according to the color scale in panel \figpanelref{d},
		measured after $N=16$ gates as a function of flux amplitude $\fluxl$ and gate separation $\bufferlength$
		(purple and black arrows in \figpanelref{b}).
		A line cut for a fixed $\fluxl$ is included in \cref{supp:la}.
		\figpanel{d}~Maximum of $P_{\qbh,2}$ (scale on the left vertical axis),
		taken over all gate separations $\bufferlength$, at each flux amplitude $\fluxl$.
		The approximate conversion to the leakage $\leakage$ of a single gate, shown on the right axis,
		is given in \cref{supp:la} (valid outside the red shaded area).
	}
	\label{fig:resonance}
\end{figure}

In order to implement the continuous gate set by means of the single parameter $\translength$
without adjusting any other parameters,
we need a fixed choice of detuning $\deltamin$ and interaction time $\pulselength$
for which the population recovery is maximal,
independently of $\translength$.
This independence relies on performing a complete population transfer to $\ket{20}$ (\cref{fig:1}h),
which requires setting the detuning to $\deltamin=0$ with high precision.
\cref{supp:param_space_sensitivity} provides a quantitative simulation of
the increase in leakage caused by a nonzero $\deltamin$
if the interaction time $\pulselength$ is kept fixed while sweeping the idle time $\translength$.

To visualize the relevant part of the parameter space,
we show the population recovery and the conditional phase
for a fixed idle phase $\kphase=\pi$ in \cref{fig:resonance}a,
which yields a qualitatively different pattern than in the case $\kphase=0$ in \cref{fig:1}e.
The central point indicated by the middle of the white arrow
is the $\cz_{0}$ gate, which is an element of the continuous gate set.
At this point,
the population recovery (shown as brightness) is insensitive to the interaction time $\pulselength$,
but depends quadratically on $\deltamin$,
see \cref{supp:param_space_sensitivity}.
This is in contrast to the $\cz_{\pi}$ gate at $\kphase=0$ (white dot in \cref{fig:1}e),
where the population recovery depends quadratically on $\pulselength$,
but is only a 4th-order function of $\deltamin$ to leading order.
Thus, whereas fixing $\kphase=0$ is adequate to resolve variations in $\pulselength$,
choosing the idle time such that $\kphase=\pi$ makes it possible
to fine-tune $\deltamin$
with a higher sensitivity than in the case $\kphase=0$.
By enabling us to establish the $\ket{11}$--$\ket{20}$ resonance more accurately,
a calibration at $\kphase=\pi$ therefore reduces the overall leakage
across the continuous gate set.

Even with this higher sensitivity,
resolving small variations in $\deltamin$
is ultimately limited by the readout fidelity and the statistical errors due to finite sampling
when measuring the population leaked to $\ket{20}$.
To overcome these limitations,
we coherently accumulate leakage
by repeating the gate $N$ times,
before measuring the population $P_{\qbh,2}$
of the second excited state of $\qbh$,
see the pulse sequence in \cref{fig:resonance}b.
Between successive gates,
the leaked population acquires a phase because it is detuned from $\ket{11}$ by $\deltamax/2\pi\approx\SI{2}{\giga \hertz}$.
This results in an interference condition that is periodic in the gate separation $\bufferlength$
with period $2\pi/\deltamax\approx\SI{0.5}{\nano \second}$,
as seen along the vertical direction in \cref{fig:resonance}c.
In addition, we sweep the pulse amplitude $\fluxl$ (horizontal axis),
which determines the detuning $\deltamin$ during the interaction.
For each value of $\fluxl$,
we then take the maximum of $P_{\qbh,2}$ over $\bufferlength$,
see \cref{fig:resonance}d.
This yields
a high-contrast proxy for the leakage of one gate,
which we use to calibrate the continuous gate set in \cref{section:results}.

On the right vertical axis of \cref{fig:resonance}d,
we estimate the leakage $\leakage$ per gate corresponding to the observed $\ket{2}$-state population,
using the model in \cref{supp:la}.
This conversion is not valid in the red shaded area,
where the leakage per gate becomes so high that the coherently accumulated leakage
corresponds to a rotation by more than $\pi$ on the $\ket{11}$--$\ket{20}$ Bloch sphere,
so that the final $\ket{2}$-state population starts to decrease.

\section{Gate set calibration and benchmarking} \label{section:results}

\begin{figure}
	\centering
	\includegraphics[width=\linewidth]{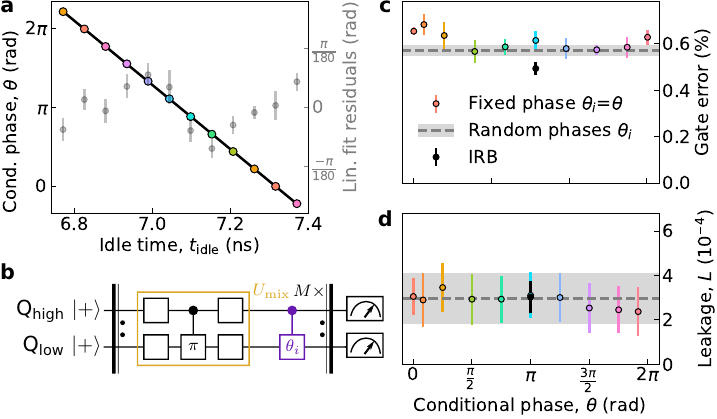}
	\caption{
		\figtitle{Gate set calibration and benchmarking.}
		\figpanel{a}~Conditional phase calibration as a function of $\translength$ (colored as in \cref{fig:1}), measured using the method detailed in \cref{supp:calib}, and linear fit (black line) with residuals (gray).
		\figpanel{b}~XEB quantum circuit, consisting of $M$ successive cycles.
		Each cycle $i$ comprises a $\cz_{\cphasei}$ gate with phase $\cphasei$, and a mixing unitary composed of randomly chosen single-qubit gates (squares) and a $\cz_{\pi}$ gate.
		\figpanel{c}~Gate error from XEB measurements, for 1750 random circuits of up to $M=256$ cycles,
		for a fixed phase $\cphasei=\cphase$ across all cycles (colored as in \cref{fig:1}),
		and for a random phase $\cphasei$ in each cycle (dashed gray line).
		The result of interleaved randomized benchmarking (IRB) is shown in black.
		\figpanel{d}~Leakage extracted from XEB and IRB measurements (legend as in \figpanelref{c}).
	}
	\label{fig:benchmarking}
\end{figure}

Based on the concepts introduced in the two previous sections,
we calibrate the continuous set of $\cz_{\cphase}$ gates,
as summarized here and detailed in \cref{supp:calib}.
We perform measurements of the population recovery and conditional phase,
as introduced in Section~\ref{section:concept}.
In these measurements, we successively sweep the pulse amplitude $\fluxl$,
the pulse length $\pulselength$ and the idle time $\translength$,
to obtain an approximate calibration of a $\cz_{\cphase}$ gate with $\cphase=0$ (i.e. $\kphase=\pi$).
For this choice of $\cphase$
the leakage is maximally sensitive to the detuning $\deltamin$,
see Section~\ref{section:resonance}.
We fine-tune $\deltamin$ 
by updating $\fluxl$ to the value that minimizes the amplified leakage after $N=16$ gates, see \cref{fig:resonance}d.
Then, we modify the idle time $\translength$ to obtain $\cphase=\pi$,
so that the leakage becomes sensitive to the interaction time $\pulselength$.
We calibrate $\pulselength$ by measuring the leakage after a single gate.

Having fixed $\fluxl$ and $\pulselength$,
we measure the conditional phase $\cphase$ for twelve uniformly spaced choices of the idle time $\translength$,
see \cref{fig:benchmarking}a.
We fit a linear model,
which is shown as a black line in \cref{fig:benchmarking}a,
and we find fit~residuals below $\pm \pi/180\;\SI{}{\radian}$,
as shown in gray in \cref{fig:benchmarking}a.
This confirms that we have calibrated a continuous gate set that behaves in close agreement with
the linear model introduced in Section~\ref{section:concept}.
The fit residuals might be caused by signal distortions in the flux control lines.
The slope of the fitted line is given by the maximal detuning
during the idle time,
$\deltamax/2\pi\approx\SI{2}{\giga\hertz}$.
With this choice of $\deltamax$, the idle time $\translength$ only needs to vary continuously
over a range of $\SI{0.5}{\nano\second}$ to cover the whole range of conditional phases.

In \cref{supp:calib}, we describe the calibration procedure
of the continuous gate set in more detail,
including calibrating a virtual $Z$ rotation \cite{McKay2017}
to correct the rotating frame by compensating the dynamic phases acquired due to the
frequency excursions of each qubit.
This yields an effective evolution
only in the $\ket{11}$--$\ket{20}$ subspace,
see \cref{eq:Hmain}.
Importantly,
changing the idle time $\translength$ does not affect the time integral of the frequency excursions (\cref{fig:1}c),
so this virtual $Z$ rotation only needs to be calibrated once for the entire gate set.

We characterize the continuous gate set using cross-entropy benchmarking (XEB) \cite{Boixo2018, Arute2019, Helsen2022a}
with a modified cycle design, overcoming limitations of existing benchmarking methods \cite{Arute2019, Hill2021, Abrams2020, Sheldon2015, Marxer2022},
see \cref{supp:xeb} for a summary of these methods.
By comparing the measured and simulated output distributions of random quantum circuits,
XEB estimates the error per quantum gate,
based on the assumption
that the distribution over circuits forms a 2-design,
meaning that they generate states
approximating aspects of the uniform distribution in the Hilbert space \cite{Boixo2018}.
The original XEB implementation \cite{Boixo2018, Arute2019}
fulfilled this condition with a random mixing unitary $\umix$ composed of single-qubit gates,
followed by the entangling two-qubit gate under test.
However, when followed by a weakly entangling $\cz_{\cphase}$ gate with $\cphase$ close to 0,
a single-qubit mixing unitary does not fulfill the randomization assumption for a two-qubit Hilbert space.
Therefore, to benchmark the whole continuous gate set,
we append the gate under test to an entangling mixing unitary $\umix$
composed of a $\cz_{\pi}$ gate and random single-qubit gates ($X_{\pi/2}$,
$Y_{\pi/2}$, $Z_{\pi/4}$),
as shown in \cref{fig:benchmarking}b.
In addition,
we perform a reference measurement where the XEB cycle consists only of the mixing unitary.
We then extract the error of the gate under test from the results of both measurements,
see \cref{supp:xeb}.
Note that $\umix$ fully explores the Hilbert space,
ensuring that the reference measurement captures possibly correlated errors during the single-qubit gates.

We extract gate errors
by executing 1750 random circuits consisting of up to $M=256$ cycles,
see \cref{fig:benchmarking}b.
By using a fixed conditional phase $\cphasei=\cphase$ in all cycles $i$,
we characterize individual $\cz_{\cphase}$ gates
for $\cphase\in\{ 0,\pi/4, \dots, 7\pi/4\}$,
as well as for weakly entangling gates with $\cphase\in\{\pm\pi/12\}$,
see the colored dots in \cref{fig:benchmarking}c.
In addition, to estimate the average error over the whole continuous gate set,
we perform XEB with a different random phase $\cphasei\in[0, 2\pi)$ in each cycle,
see the dashed line in \cref{fig:benchmarking}c.
Each individual gate yields a similar error,
with a mean of $\SI{0.61}{\%}$ in good agreement with the average measurement at $\SI{0.57\pm0.02}{\%}$,
and a maximal deviation from the mean at $\SI{0.68\pm0.04}{\%}$ for $\cphase=\pi/12$.
These gate errors are similar to the infidelity of $\SI{0.7}{\%}$ reported for $\cphase=\pi$
in the continuous set of $\cz_{\cphase}$ gates in \cite{Barends2015},
but we note that this gate set was limited to the interval $\cphase/\SI{}{\radian}\in[0.7, 4.0]$.
Moreover, the gate errors reported above are lower than the median error of $\SI{2.7}{\%}$
reported for gates from the $XY$ family in \cite{Abrams2020}
and of $\SI{1.7}{\%}$ for a $\cz_{\pi}$ gate within the $\cz_{\cphase}$ gate set in \cite{Lacroix2020}.
At the expense of the additional overhead of an implementation with tunable couplers,
a lower average infidelity of $\SI{0.2}{\%}$ was achieved for a continuous set of $\cz_{\cphase}$ gates in \cite{Foxen2020}.
For discrete $\cz_{\pi}$ gates implemented with fixed coupling based on a similar gate concept as presented here,
a similar infidelity as in our experiment was reported for the best gate in \cite{Krinner2022} ($\SI{0.6}{\%}$)
and for the best gate in \cite{Negirneac2021} ($\SI{0.5}{\%}$).

For a possible explanation of the variations of the gate error
as a function of $\cphase$ observed in \cref{fig:benchmarking}c,
consider that the qubits typically cross two-level defects \cite{Wang2015n, Mueller2019, ColaoZanuz2024}
during their frequency excursions.
Interference effects resulting from these crossings \cite{Ivakhnenko2023}
depend on the idle time $\translength$
and are not taken into account in the calibration.
We estimate the purity of the elements of the continuous gate set using speckle purity benchmarking \cite{Arute2019},
and obtain incoherent errors that are similar to the total gate errors
within error bars,
indicating that control errors do not provide a significant contribution to the total gate error.
We also perform interleaved randomized benchmarking (IRB) \cite{Magesan2012a, Magesan2012} for the Clifford gate $\cz_{\pi}$,
see the black dot in \cref{fig:benchmarking}c.
The gate error of $\SI{0.49\pm0.03}{\%}$ reported by IRB is in reasonable agreement with the XEB results.

Finally, we extract leakage from XEB \cite{Wood2017, Lazar2023a} and IRB \cite{Chen2016, Lazar2023a}, as shown in \cref{fig:benchmarking}d.
The observed values all lie between $2\times 10^{-4}$ and $4\times 10^{-4}$,
indicating uniformly low leakage over the entire gate set.
This is slightly lower than the lowest leakage values reported for gates
from the continuous gate sets presented in \cite{Barends2015, Lacroix2020, Foxen2020, Collodo2020},
which lie between $5\times 10^{-4}$ and $15\times 10^{-4}$.
A similar amount of leakage was also found for the discrete $\cz$ gates implemented in \cite{Negirneac2021, Krinner2022}.

\section{Conclusion} \label{section:conclusion}

By extending the net-zero $\cz_{\pi}$ gate used in previous work \cite{Negirneac2021, Herrmann2022, Krinner2022},
we have designed a continuous set of $\cz_{\cphase}$ gates
in a fixed-coupling architecture.
We have shown how fine-tuning the $\ket{11}$--$\ket{20}$ resonance
yields a parameter space with a simple structure,
where the conditional phase is the sum of a constant geometrical phase
and a controllable dynamic phase.
As a result, the conditional phase has a linear dependence on a single control parameter,
the idle time $\translength$.
In addition, the gate set achieves low leakage and high fidelity independently of $\translength$.
In contrast, other schemes such as \cite{Barends2015, Lacroix2020, Foxen2020}
must keep track of several interrelated parameters to define a continuous gate set.

In order to calibrate the $\ket{11}$--$\ket{20}$ resonance with sufficient precision,
we have introduced a coherent leakage amplification method,
which allows measuring leakage below the readout limit.
We expect that this method can contribute
to both the calibration and benchmarking procedures
in broader contexts where low-leakage gates are a key requirement,
including two-qubit gates for quantum error correction \cite{Aliferis2007b, Fowler2013, Ghosh2015a, Suchara2015, Bultink2020, Varbanov2020, Battistel2021}.
We have also introduced a modified XEB cycle design, capable of benchmarking weakly entangling gates
by interleaving them with an entangling mixing unitary.

We have implemented and characterized the continuous gate set
in an experiment using two qubits with fixed coupling on a six-transmon quantum device,
demonstrating state-of-the-art fidelities and leakage across the entire gate set.
As the benchmarking results indicate that incoherent errors are dominant in our implementation,
we expect that the gate errors can be further reduced
by deploying the continuous gate set on qubits with longer coherence times.

\section*{Author contributions}
C.H. and C.S. conceived the experiments,
C.S. and K.H. performed the measurements,
and C.S. analyzed the data.
C.S., K.H. and A.R. performed numerical simulations,
and C.S., S.L. and N.L. contributed to the measurement software.
A.R. and C.S. designed the device,
and D.C.Z. and A.F. fabricated it.
C.H. and A.W. supervised the work.
C.S. and C.H. wrote the manuscript with input from all co-authors.

\section*{Acknowledgments}

We kindly thank Ilya Besedin, Quentin Ficheux,
Christian Kraglund Andersen and Christopher Eichler for insightful discussions,
and Johannes Kn\"orzer and Boris Varbanov for feedback on the manuscript.
Research was sponsored by IARPA and the Army Research Office,
under the Entangled Logical Qubits program,
and was accomplished under Cooperative Agreement Number W911NF-23-2-0212,
by the EU program H2020-FETOPEN project 828826 Quromorphic,
by the Baumgarten foundation
and by ETH Zurich.
The views and conclusions contained in this document are those of the authors
and should not be interpreted as representing the official policies,
either expressed or implied, of IARPA,
the Army Research Office, or the U.S. Government.
The U.S. Government is authorized to reproduce and distribute reprints
for Government purposes notwithstanding any copyright notation herein.

\appendix
\crefalias{section}{appendix}

\section{Experimental setup and quantum device} \label{supp:setup}

\begin{figure}
	\centering
	\includegraphics[width=\linewidth]{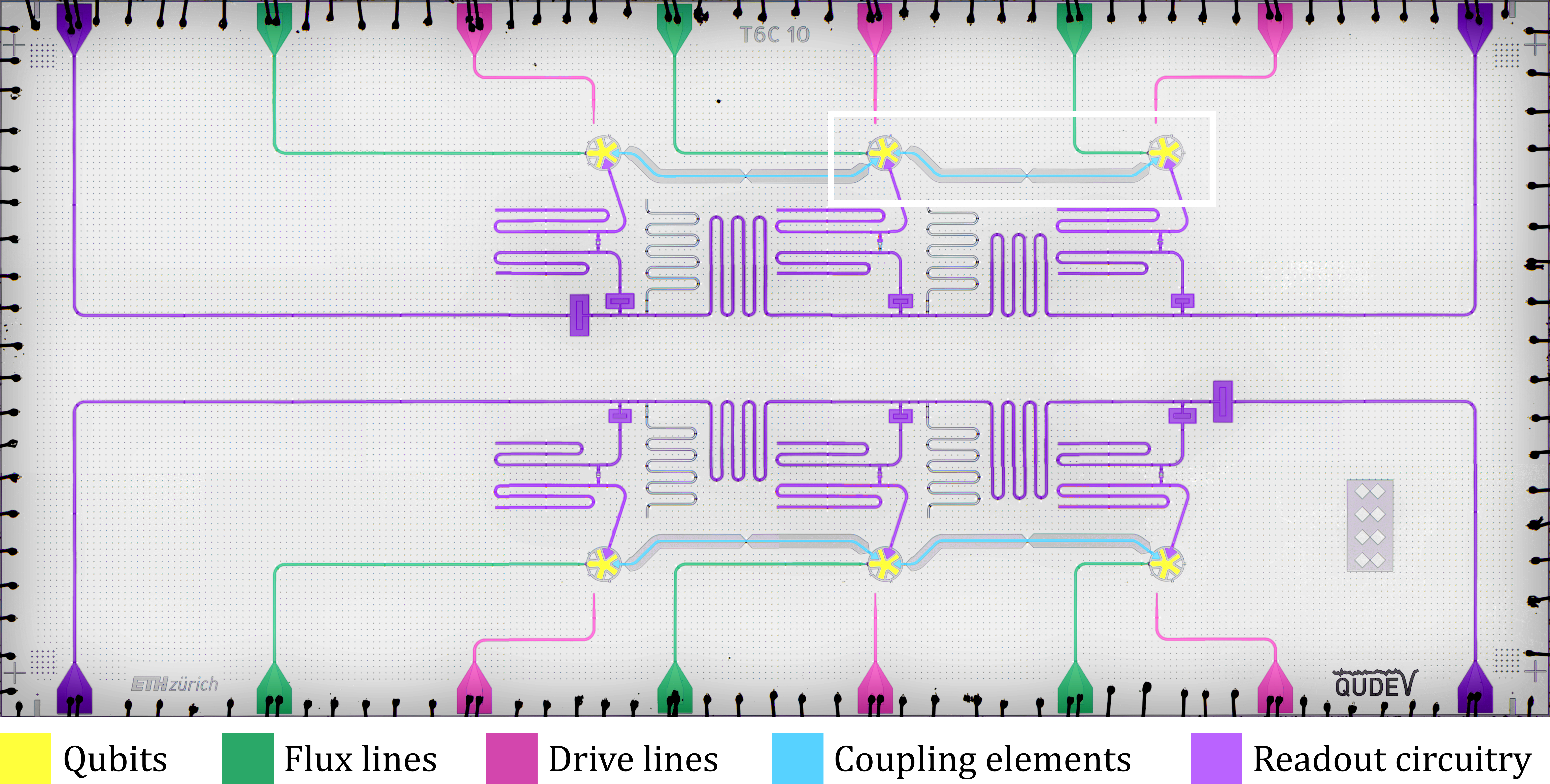}
	\caption{
		\figtitle{False-color micrograph} of the device used in a preliminary version of this experiment; the device used in the final experiment is nominally identical.
		Each transmon qubit consists of a capacitive island (yellow) connected to the ground plane via a superconducting quantum interference device (SQUID),
		and capacitively coupled to a drive line (pink) and the readout circuitry (purple).
		Fast flux control is provided by individual flux lines (green), which are inductively coupled to the SQUID of each qubit.
		The capacitive qubit-qubit coupling is implemented with fixed-frequency coupling resonators (light blue).
		The measurements have been performed on the pair of qubits indicated by a white frame.}
	\label{fig:supp_chip}
\end{figure}

\begin{figure}
	\centering
	\includegraphics[width=\linewidth]{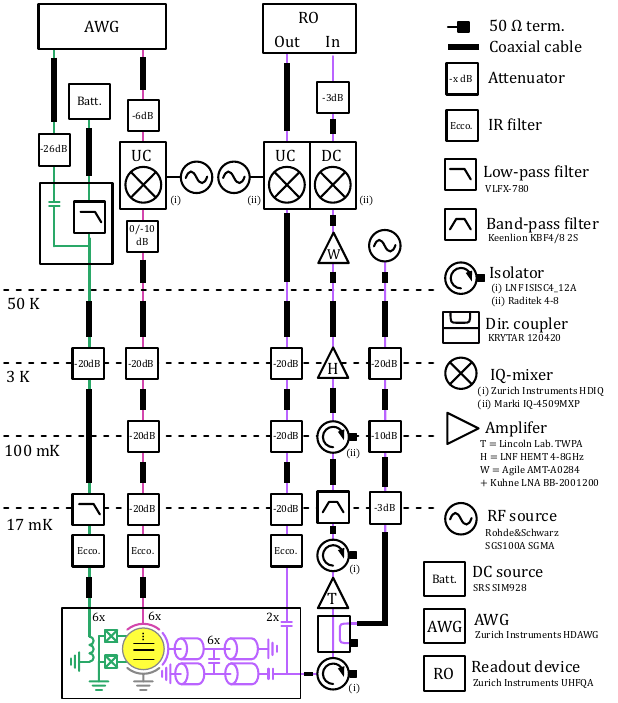}
	\caption{
		\figtitle{Simplified schematic of the experimental setup},
		including the quantum device.
		Coupling elements between transmons are not represented. See the text for details.}
	\label{fig:supp_setup}
\end{figure}

We performed the experiment on a pair of transmon qubits,
on a superconducting quantum device hosting two chains of three qubits,
see micrograph in \cref{fig:supp_chip}.
We have fabricated the device according to the methods specified in \cite{ColaoZanuz2024} and summarized below.
We have used a high-resistivity intrinsic silicon substrate, with a niobium film as the base metallization layer.
The circuit elements are defined using photolithography and reactive-ion etching.
The different regions of the ground plane are electrically connected using aluminum/titanium/aluminum trilayer airbridges.
We have deposited the aluminum/aluminum-oxide/aluminum Josephson junctions using double-angle shadow evaporation,
with features patterned by electron-beam lithography.
We have exposed and deposited aluminum bandages \cite{Dunsworth2017} to ensure contact between the leads of the Josephson junctions and the base metallization.
The device is mounted in a magnetically-shielded sample holder,
and operated below $\SI{20}{\milli\kelvin}$ in a dilution refrigerator, see the setup schematic in \cref{fig:supp_setup}.

The two qubits used in this work are read out via dedicated readout resonators and Purcell filters,
coupled to a common feedline with frequency multiplexing \cite{Heinsoo2018}.
The readout signal is amplified by a traveling wave parametric amplifier (TWPA) \cite{Macklin2015},
followed by a cryogenic high-electron mobility transistor (HEMT) amplifier, and two low-noise amplifiers at room temperature.
We initialize the qubits in their ground states using a method based on real-time feedback,
by repeatedly performing three-level readout followed by single-qubit drive pulses ($\ket{2}\leftrightarrow\ket{1}$ and $\ket{1}\leftrightarrow\ket{0}$) conditioned on the readout results \cite{Riste2012b, Salathe2018}.

We implement single-qubit gates by applying microwave pulses via dedicated drive lines,
which are capacitively coupled to the qubits.
The pulses are generated by an arbitrary waveform generator and up-converted by in-phase \& quadrature (IQ) mixers.
Single-qubit $X$ and $Y$ gates are realized as derivative removal by adiabatic gates (DRAG) \cite{Motzoi2009, Lazar2023}.
Single-qubit $Z$ gates are implemented as so-called virtual gates by adjusting the phase of subsequent $X$ and $Y$ pulses \cite{McKay2017}.

The transition frequency of each qubit is controlled by the current flowing in an inductively-coupled flux line,
which sets the magnetic flux threading the SQUID loop of the qubit.
The idling transition frequency is set by a constant current from an isolated voltage source.
The frequency excursions implementing the two-qubit gates are realized by fast flux pulses,
generated by an arbitrary waveform generator.
The pulses are combined with the bias current via a custom-made bias-tee.
See \cref{supp:waveform} for more details on flux pulse control.
We fit a coupled transmon-resonator model to the transition frequency
of each qubit as a function of the flux threading its SQUID loop.
These models are represented in \cref{fig:1}b,
and used in the experiments to determine the conversion between flux pulse amplitudes and qubit frequencies.

The coupling between neighboring qubits is mediated by a fixed-frequency resonator (light blue in \cref{fig:supp_chip}).
Both qubits ($\qbh$, $\qbl$) are capacitively coupled to the resonator, with coupling rates $g_{\subsqbh}$ and $g_{\subsqbl}$, respectively.
The resonator is designed to have a resonance frequency of $\SI{24}{\giga\hertz}$,
such that the frequency differences $\Delta_{\subsqbh}$ and $\Delta_{\subsqbl}$
between the resonator and the transmons are in the dispersive regime, $\Delta\gg g$.
In this regime, and with the resonator initially in the vacuum state, the effect of the resonator reduces to an effective capacitive coupling \cite{Blais2021}
\begin{align}
	\jhalfo = 2J_1 = g_{\subsqbh}g_{\subsqbl}\left(\frac{1}{\Delta_{\subsqbh}}+\frac{1}{\Delta_{\subsqbl}}\right)
\end{align}
where $\jhalfo$ is the frequency splitting of the coupling Hamiltonian in the one-excitation subspace,
and $J_1$ the corresponding matrix element.
The coupling rate between $\ket{20}$ and $\ket{11}$ (two-excitation subspace) is given by $\jhalft = 2J_2 = \sqrt{2}\jhalfo$.
This coupling $\jhalft$ is the dominating relevant term in the Hamiltonian only when $\ket{20}$ and $\ket{11}$ have similar energies, within a range of approximately $\jhalft$ or less.
In this small range of qubit frequencies,
the detunings $\Delta_{\subsqbh}$ and $\Delta_{\subsqbl}$ are close to constant,
and $\jhalft$ can therefore be considered constant in the regime relevant for a two-qubit gate.

\begin{table}[]
	\begin{tabular}{|l|l|l|}
		\hline
		\textbf{Name}																& $\qbh$		&	$\qbl$		\\ \hline
		Qubit idle frequency, $\omega/2\pi$ $(\SI{}{\giga\hertz})$					& $6.278$		& $4.083$			\\ \hline
		Qubit anharmonicity, $\alpha/2\pi$ $(\SI{}{\giga\hertz})$					& $-0.167$		& $-0.191$		\\ \hline
		Readout frequency, $\omega_{\mathrm{RO}}/2\pi$ $(\SI{}{\giga\hertz})$		& $7.423$		& $7.215$			\\ \hline
		Three-state readout error, $\epsilon_{\mathrm{RO}}$ $(\%)$					& $1.49$		& $1.36$			\\ \hline
		Residual population $P_{\mathrm{res}}$ $(\%)$								& $0.02$		& $0.43$			\\ \hline
		Single-qubit RB error, $\epsilon_{\mathrm{1Q}}$ $(\%)$						& $0.07$		& $0.05$			\\ \hline
		Lifetime, $T_{\mathrm{1}}$ $(\SI{}{\micro\second})$ 						& $\SI{32.6\pm0.1}{}$	& $\SI{88.0\pm0.4}{}$	\\ \hline
		Lifetime of the $\ket{2}$ state, $T_{\mathrm{1}}^{\ket{2}}$ $(\SI{}{\micro\second})$ 	& $\SI{14.7\pm0.3}{}$	& $\SI{30.3\pm0.7}{}$	\\ \hline
		Ramsey decay time, $T^*_{\mathrm{2}}$ $(\SI{}{\micro\second})$				& $\SI{23.9\pm1.1}{}$	& $\SI{46.8\pm1.3}{}$	\\ \hline
		Echo decay time, $T^{\mathrm{E}}_{\mathrm{2}}$ $(\SI{}{\micro\second})$		& $\SI{28.9\pm0.5}{}$	& $\SI{50.3\pm1.1}{}$	\\ \hline
		Qubit-qubit coupling $J_2/2\pi$ $(\SI{}{\mega\hertz})$						& \multicolumn{2}{c|}{$\SI{11.38\pm0.01}{}$}	\\ \hline
		Residual coupling $\Delta_{ZZ}/2\pi$ $(\SI{}{\kilo\hertz})$					& $\SI{1.3\pm0.7}{}$	& $\SI{4.4\pm0.7}{}$	\\ \hline
		Waveform smoothing, $\sigma$ $(\SI{}{\nano\second})$						& $1.02$		& $1.35$			\\ \hline
	\end{tabular}
	\caption{
		\figtitle{Qubit parameters and properties.}
		See the text for details.
	}
	\label{tab:device_params}
\end{table}

\cref{tab:device_params} shows a summary of the properties of the quantum device.
The qubit-qubit coupling is specified at the interaction frequency of the gate,
and extracted from a fit to the exchange pattern in \cref{fig:supp_calib_chevron},
see \cref{supp:calib} for a description of the measurement.
The residual couplings $\Delta_{ZZ}$ are extracted from Ramsey experiments
with the other qubit either prepared in the ground or excited state.
The error bars are standard error estimates calculated from the
variance of the fit residuals.
For the parameter $\sigma$ describing the effective Gaussian filtering of flux control waveforms, see \cref{supp:waveform}.

\section{Flux waveform generation} \label{supp:waveform}

To implement two-qubit gates, we apply flux pulses via inductively-coupled flux lines, as described in \cref{supp:setup}.
Here, we summarize the methods which we apply to characterize and compensate pulse distortions,
we comment on effects due to non-zero rise times of the flux pulses,
and we discuss the time resolution for adjusting the idle time $\kphase$.

Due to filters in the flux control lines and unwanted effects such as parasitic capacitances,
the pulses are distorted before reaching the SQUID loops of the qubits.
We characterize long-time distortions above $\SI{60}{\nano\second}$
with the spectroscopic methods described in \cite{Johnson2011PhD, Krinner2022}
and short-time distortions with the cryoscope method from \cite{Rol2020},
as detailed in \cite{Hellings2025}.
We then apply digital predistortion filters to compensate the distortions,
such that the combination of digital preprocessing and physical flux line behaves in good approximation as a Gaussian filter.
The resulting limited bandwidth makes the two-qubit gates adiabatic with respect to the transmon transition frequencies.

To characterize the width $\sigma$ of this effective Gaussian filtering,
we perform dynamic phase measurements, as described in \cref{supp:calib},
while sweeping the duration and the amplitude of a rectangular flux pulse on a single qubit.
To this dataset, we fit a numerical model
which expresses the dynamic phase of the qubit as the time integral of
the detuning from the idle frequency during the Gaussian-filtered flux pulse.
We consider the parameter $\sigma$ of the Gaussian filter as fit parameter,
yielding the values shown in \cref{tab:device_params}.
The simulations rely on the flux-frequency characteristic of the transmons
obtained in \cref{supp:setup} and shown in \cref{fig:1}b.

\begin{figure}
	\centering
	\includegraphics[width=\linewidth]{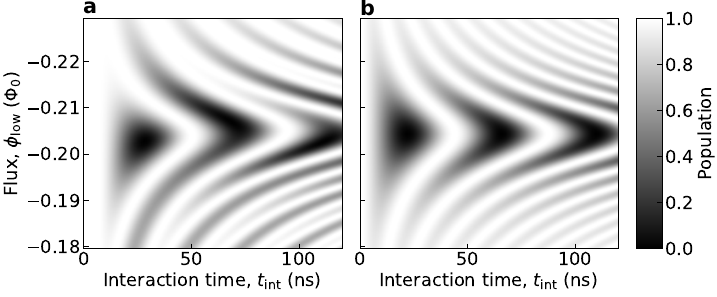}
	\caption{
		\figtitle{Simulated population recovery} after one gate, for
		\figpanel{a}~Gaussian-filtered net-zero waveforms (see the experimentally measured values of $\sigma$ in \cref{tab:device_params}) and
		\figpanel{b}~unfiltered waveforms ($\sigma=\SI{0}{\nano\second}$).
	}
	\label{fig:supp_chevrons_sigma}
\end{figure}

To study the effect of Gaussian filtering on the population recovery after a two-qubit gate,
we show simulation results in \cref{fig:supp_chevrons_sigma}.
We integrate the Schr\"odinger equation including all energy levels of the two-transmon system up to two excitations,
with a time-dependent Hamiltonian defined by the Gaussian-filtered pulse shapes,
and assuming a fixed qubit-qubit coupling (as defined in \cref{supp:setup}).
These simulations do not take into account residual crosstalk between the flux lines,
which can influence the exact flux pulse amplitude required to reach the resonance condition in the experiment
(see the different vertical axis in \cref{fig:1}e).
We observe that Gaussian-filtered net-zero waveforms yield
an irregular spacing in the curves of population recovery in \cref{fig:supp_chevrons_sigma}a,
which is in qualitative agreement with the experimental data in \cref{fig:1}e.
This effect disappears when performing the simulation with piecewise-constant waveforms
without applying the Gaussian filter, see \cref{fig:supp_chevrons_sigma}b.

\begin{figure}
	\centering
	\includegraphics[width=\linewidth]{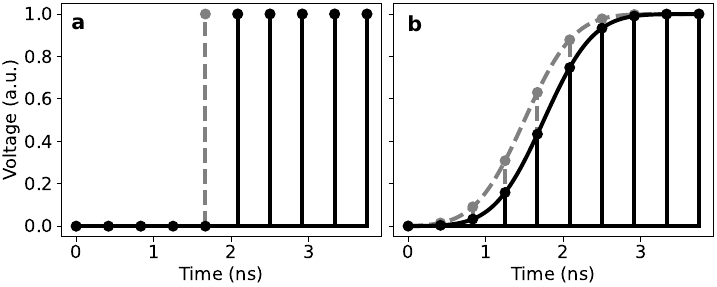}
	\caption{
		\figtitle{Concept of shifting (solid back) a given waveform (dashed gray)}
		by means of adjusting the amplitude of the points during the pulse edges.
		The waveform samples are shown on a $1/(\SI{2.4}{\giga\hertz})$ grid,
		as produced by the AWG employed in the experiment.
		\figpanel{a}~Shifting a waveform with a sharp rising edge, by shifting all samples by a sampling period.
		\figpanel{b}~Shifting a Gaussian-filtered waveform ($\sigma=\SI{0.5}{\nano\second}$) by half the sampling period,
		by adjusting the amplitudes of the samples during the smooth rising edge.
	}
	\label{fig:supp_subsample_sketch}
\end{figure}

Setting the conditional phase $\cphase$ of a $\cz_{\cphase}$ gate by varying the idle time $\translength$ in the middle of the waveforms,
as described in \cref{section:concept},
requires to adjust $\translength$ with a resolution below the sampling period
$T_{\mathrm{S}}=\SI{0.42}{\nano\second}$ of the AWG.
We first apply a Gaussian filter of width $\sigma=\SI{0.5}{\nano\second}$,
then adjust the amplitudes of the samples at the pulse edges to account for appropriate time shifts as illustrated in \cref{fig:supp_subsample_sketch}, and then apply all other digital filtering.
A similar concept was used e.g. in \cite{Kong2023}.
To obtain an order-of-magnitude estimate of the time precision that can be reached,
we approximate a rising edge as a line with constant slope
within a rise time of $t_{\mathrm{rise}}\approx2\sigma=\SI{1}{\nano\second}$.
With a maximum amplitude $A_{\mathrm{max}}=\SI{1}{\volt}$
and an amplitude precision $\delta_A=\SI{0.1}{\milli\volt}$ (from the datasheet of the AWG),
we expect a timing precision $\delta_t\approx t_{\mathrm{rise}}\delta_A/A_{\mathrm{max}} \approx \SI{1}{\nano\second}\times\SI{0.1}{\milli\volt}/\SI{1}{\volt} = \SI{0.1}{\pico\second}$.

When calibrating the continuous gate set, we observe residuals below $\pm 1^{\circ}$ in the conditional phase (\cref{fig:benchmarking}a),
corresponding to a required timing precision of $(1^{\circ}/360^{\circ})/\deltamax=\SI{1.3}{\pico\second}\gg\delta_t$
for a maximal detuning of $\deltamax = \SI{2.0}{\giga\hertz}$.
This suggests that the time resolution $\delta_t$ is not the limiting factor in our experiment.

\section{Calibration procedure} \label{supp:calib}

Implementing the continuous set of $\cz_{\cphase}$ gates requires calibrating two basic operations:
the half-waveforms (responsible for the population exchange), and the idle time between them (responsible for the idle phase).

We initialize all the two-qubit gate parameters by estimating them from the flux dependence of the transition frequencies
obtained in \cref{supp:setup} and shown in \cref{fig:1}b.
Applying flux pulses to both qubits gives the degree of freedom
of choosing the frequency at which the qubits interact during the gate \cite{Herrmann2022}.
When selecting this interaction frequency,
we avoid collisions with two-level defect modes by monitoring the population loss
of the excited state of each qubit during a flux pulse of variable amplitude \cite{Shalibo2010, Lisenfeld2016}.
We initialize the interaction time $\pulselength$ based on the coupling strength from the device design.
In all measurements, we add $\SI{10}{\nano\second}$ buffers before and after the flux pulses,
to prevent that possible residual control pulse distortions influence the next gate.

\begin{figure}
	\centering
	\includegraphics[width=\linewidth]{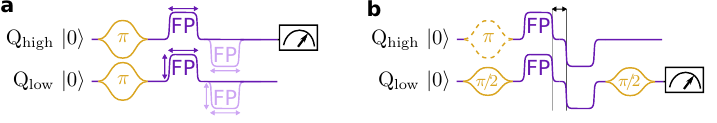}
	\caption{
		\figtitle{Calibration measurements.}
		\figpanel{a}~Pulse sequence to estimate the state populations,
		by reading out (black box) the high-frequency qubit
		after applying half-waveforms (dark purple) or full waveforms (shaded)
		of flux pulses (FP)
		to an initial $\ket{11}$ state prepared by  $\pi$ pulses (yellow).
		Horizontal arrows indicate a sweep of the interaction time
		and vertical arrows a sweep of the flux at the low-frequency qubit.
		\figpanel{b}~Pulse sequence to estimate the conditional phase of $\ket{11}$,
		extracted from a Ramsey-type measurement on the low-frequency qubit,
		with the high-frequency qubit prepared in $\ket{1}$ (dashed $\pi$-pulse) or $\ket{0}$,
		and sweeping the phase of the final $\pi/2$-pulse.
	}
	\label{fig:supp_calib_pulse}
\end{figure}

\begin{figure}
	\centering
	\includegraphics[width=\linewidth]{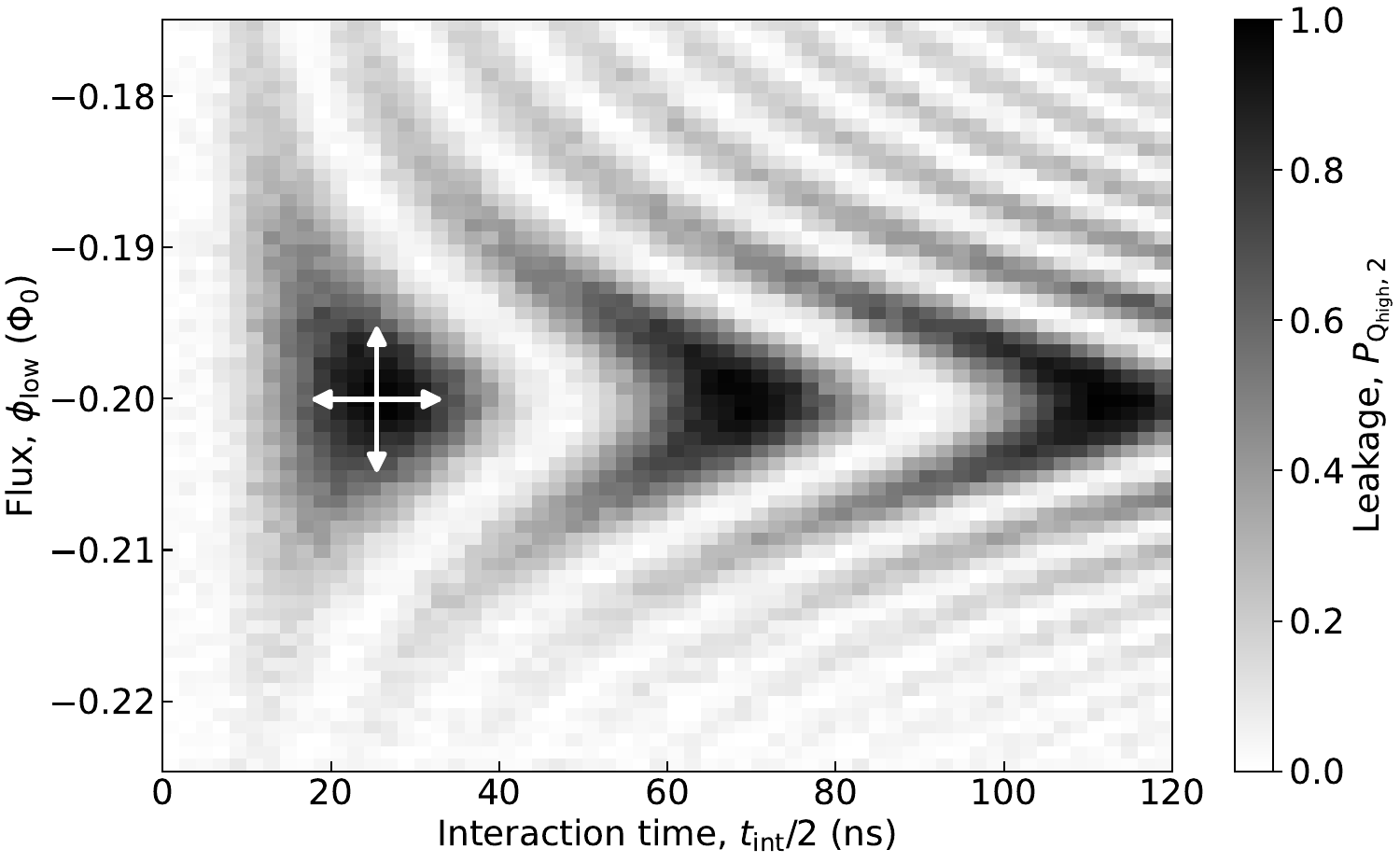}
	\caption{
		\figtitle{Measured population recovery after a rectangular half-waveform,}
		as a function of the interaction time $\pulselength$ and the flux $\fluxl$ at the low-frequency qubit during the interaction.
		White arrows around the point of maximal $\ket{20}$ population illustrate the sweeps shown in \cref{fig:supp_calib_pulse}a.
	}
	\label{fig:supp_calib_chevron}
\end{figure}

We start by calibrating a rectangular pulse for each qubit,
to perform a full exchange to $\ket{20}$.
We use the pulse sequence shown in \cref{fig:supp_calib_pulse}a to measure the populations after this waveform (dark purple).
When performing three-level readout,
we extract the state populations from the measured signal quadratures
by orthogonal projection on calibration points \cite{Magnard2018}.
We measure the excited state populations $P_{\qbh,1}$ and $P_{\qbh,2}$ of $\qbh$
as a proxy for the population of $\ket{11}$ and the leakage to $\ket{20}$.
The resulting characteristic pattern of population exchange is shown in \cref{fig:supp_calib_chevron},
and can be measured at low resolution at the beginning of the calibration to check the initial parameters.
To obtain a full population exchange,
we calibrate the duration and amplitude of the rectangular pulses
by iterating one-dimensional sweeps (white arrows)
and maximizing quadratic fits of the leakage.

We then construct a $\cz_{\cphase}$ gate
by concatenating two such rectangular pulses with opposite signs
and otherwise identical pulse parameters (\cref{fig:1}a).
We allow a variable idle time $\translength$ between the two rectangular pulses.
At this point, we can directly obtain an approximate calibration of the continuous gate set,
by using the Ramsey-type pulse sequence shown in \cref{fig:supp_calib_pulse}b
to measure the conditional phase as a function of $\translength$.
The calibration presented up to this point routinely achieved gate errors below $\SI{1}{\%}$ on the considered pair of qubits.
In order to reproducibly reach the low residuals shown in \cref{fig:benchmarking}a as well as low leakage, we now introduce additional fine-tuning steps.

We fine-tune the parameters, $\deltamin$ and $\pulselength$,
around the $\cz_{0}$ and $\cz_{\pi}$ gate respectively,
where leakage becomes a quadratic function of the considered parameter,
see the Taylor expansions in \cref{supp:param_space_sensitivity}.
We first set a value of $\translength$ that yields the conditional phase $\cphase=0$,
and fine-tune $\deltamin$ by choosing the flux pulse amplitude $\fluxl$ which minimizes the measured accumulated leakage after 16 gates (\cref{fig:resonance}d).
We then set $\translength$ to realize a $\cz_{\pi}$ gate,
and fine-tune $\pulselength$ by minimizing the measured leakage after one gate.

Having fine-tuned the pulse parameters $\fluxl$ and $\pulselength$,
we characterize the conditional phase $\cphase$ as a function of $\translength$, as shown in \cref{fig:benchmarking}a.
To implement any $\cz_{\cphase}$,
we then pick $\translength$ from this dataset,
either using a linear fit, or an interpolation for robustness to residual inaccuracies in the calibration of other parameters.
In the datasets shown in this paper, we use a cubic spline interpolation.
When applying gates in a quantum circuit,
we keep their duration fixed by adjusting the buffers around the flux pulses
to compensate for the variable idle time between the two half-waveforms.

\begin{figure}
	\centering
	\includegraphics[width=\linewidth]{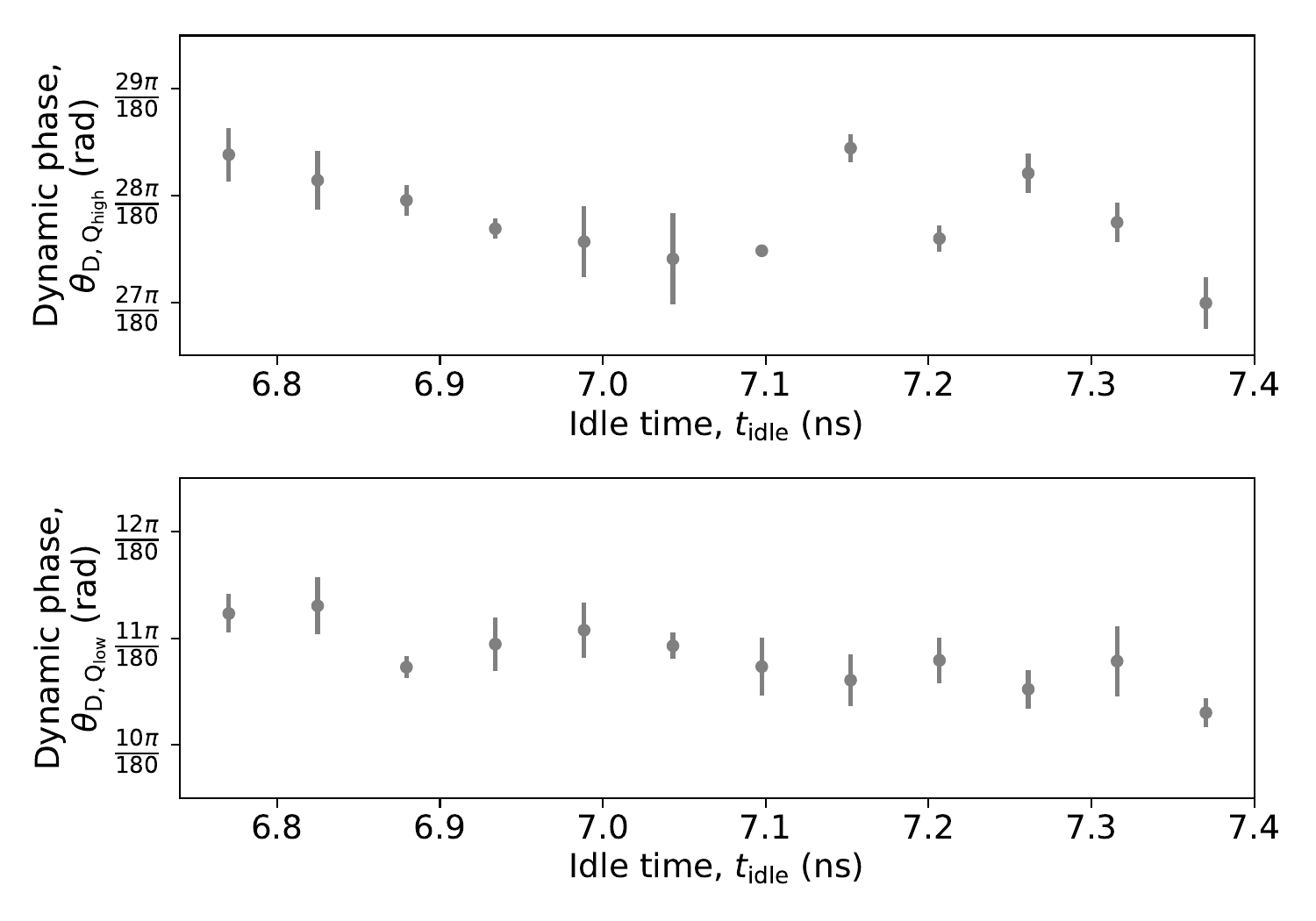}
	\caption{
		\figtitle{Dynamic phase}
		measured as a function of the idle time $\translength$.
	}
	\label{fig:supp_dynphase_sweep}
\end{figure}

At the end of the calibration procedure, we characterize the single-qubit dynamic phases on $\ket{10}$ and $\ket{01}$,
using a Ramsey-type measurement \cite{DiCarlo2009}.
The dynamic phases depend on the shapes of the half-waveforms, but not on the variable idle time $\translength$ between them,
so that a single calibration for the entire gate set is sufficient.
We confirm this experimentally by performing the measurements as a function of $\translength$ shown in \cref{fig:supp_dynphase_sweep}.
The residual variations below $\pm \pi/180\;\SI{}{\radian}$ may be due to residual distortions in the flux control lines.
The constant dynamic phase is a significant simplification compared to the continuous gate set presented in [10],
where the dynamic phases strongly varied for different values of the conditional phase $\cphase$.
We cancel the dynamic phases using so-called virtual $Z$ gates \cite{McKay2017},
i.e., by a re-definition of the rotating frame.
An alternative would be to perform the cancellation using additional low-amplitude flux pulses, as done in \cite{Negirneac2021}.

Finally, we note that other parameterizations of the idle phase are possible.
For example, one can use a non-zero flux amplitude during the idle time,
changing the integral of the detuning during this time and therefore the conditional phase.
Due to the non-linear flux-frequency relation of the transmons and the finite slope of the pulse edges,
this, however, does not yield a linear relation
between the control parameter (flux amplitude) and the conditional phase.
Moreover, due to Gaussian filtering of the flux control pulses,
changing the amplitude during the idle time has the side effect of distorting the edges of the main half-pulses,
effectively changing their duration, the interaction time $\pulselength$.
Since this turns the interaction time into a function of the parameter that controls the conditional phase,
the continuous gate set cannot be controlled with a single parameter in this parameterization.
A variant of this approach was pursued in \cite{Negirneac2021}
by smoothly adjusting the amplitude of a single point at the edge of each half-waveform.
When not only calibrating a single $\cz_{\pi}$ gate, as in \cite{Negirneac2021}, but an entire continuous gate set,
this variant would have the same drawbacks as described above for the approach of adjusting the amplitude during the idle time.

\section{Geometrical phase} \label{supp:geom_phase}

\begin{figure}
	\centering
	\includegraphics[width=0.8\linewidth]{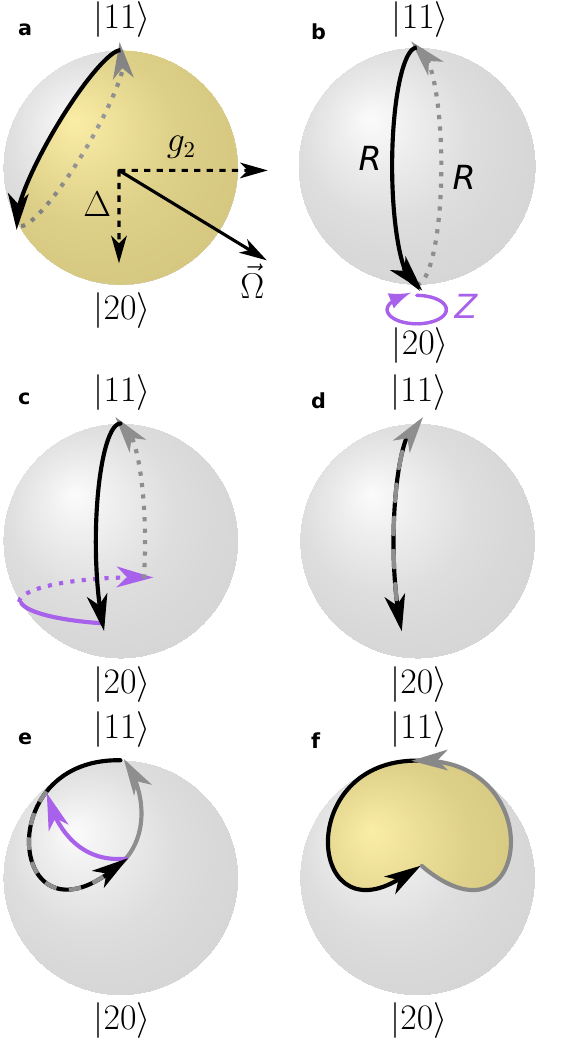}
	\caption{
		\figtitle{State trajectories and geometrical phase} during a net-zero pulse.
		The solid angle (yellow) is equivalent to the conditional phase
		in the absence of a $Z$ rotation (purple), see the text for details.
		We use the convention that a positive phase increase of $\ket{20}$
		corresponds to a counter-clockwise rotation when seen from the top of the Bloch sphere.
		Accordingly, a $Z$ rotation rotates clockwise,
		for a detuning $\Delta>0$ pointing downward.
		The time evolution is represented for
		\figpanel{a}~a pulse with $\kphase=0$ as in \cref{fig:1}g,
		\figpanel{b}~a $\cz_{\cphase}$ gate from the continuous gate set,
		\figpanel{c}~a pulse implementing a short $\cz_{0}$ gate and
		\figpanel{d}~the unitary equivalent to \figpanelref{c} after commuting out the $Z$ rotation
		as explained in the text,
		yielding a vanishing solid angle,
		\figpanel{e}~a pulse with freely chosen parameters and
		\figpanel{f}~the unitary equivalent to \figpanelref{e} after commuting out the $Z$ rotation.
	}
	\label{fig:supp_geometrical_phase}
\end{figure}

We express the conditional phase $\cphase$
of a controlled arbitrary-phase gate
\begin{align}
	\label{eq:CZphi}
	\cz_{\cphase} =& \begin{pmatrix}
		1&0&0&0\\
		0&1&0&0\\
		0&0&1&0\\
		0&0&0&e^{i\cphase}
	\end{pmatrix}
\end{align}
as a function of the state trajectory on the $\ket{11}$--$\ket{20}$ Bloch sphere.
We consider the implementation based on net-zero pulses presented in the main text and summarized in \cref{fig:1}.
The time evolution during the net-zero pulses can be seen as equivalent to a Mach-Zehnder interferometer
with two unbalanced beamsplitters $R$
and an arbitrary phase-shifter $Z$ \cite{Negirneac2021}.
The presence of the phase-shifter makes it challenging to gain intuition about the conditional phase.
We thus instead interpret the state evolution in a geometric picture,
where the phase at the end of a cyclic trajectory is given by
the Aharonov-Anandan phase \cite{Aharonov1987}:
\begin{align}
	\cphase
	&=\underbrace{\oint \bra{\Psi}\frac{d}{\diff t}\ket{\Psi} \diff t}_{\cphase_{\mathrm{G}}}
	\underbrace{- \oint \bra{\Psi}H\ket{\Psi} \diff t}_{\cphase_{\mathrm{D}}}\label{eq:geom_phase_nonadiab}.
\end{align}
In the special case of an adiabatic evolution where the state stays aligned with the axis of rotation at all times,
$\theta$ would be fully determined by the geometrical phase (Berry phase),
$\cphase_{\mathrm{G}}=-S/2$,
where $S$ is the solid angle enclosed by the trajectory on the Bloch sphere \cite{Berry1984}.
For a non-adiabatic evolution,
\cref{eq:geom_phase_nonadiab} also includes the dynamic phase contribution $\cphase_{\mathrm{D}}$
from the instantaneous energy of the state.

From \cref{eq:Hmain}, we can rewrite the time-dependent Hamiltonian restricted to the two relevant states as
\begin{align}
	\label{eq:Hsupp}
	H(t) = \begin{pmatrix}
		0&\jhalft/2\\
		\jhalft/2&\deltat\\
	\end{pmatrix}
	= \deltat \frac{\sigma_e}{2} + \jhalft \frac{\sigma_x}{2}
\end{align}
where
$I$, $\sigma_z$, $\sigma_x$ are the identity and Pauli matrices
acting on the $\ket{11}$--$\ket{20}$ subspace,
and we have defined $\sigma_e = I-\sigma_z$.
This yields an instantaneous axis of rotation on the Bloch sphere $\vec{\Omega} = (\jhalft,0,-\deltat)$.
Assuming that $\deltat$ is a piecewise-constant, symmetric function of time suitable to reach full population recovery,
a net-zero pulse sequence implements the unitary
\begin{align}
	\label{eq:geom_parameterization}
	U = R_{\lambda}^{\vec{r}}Z_{\cphase_{\mathrm{D}}}R_{\lambda}^{\vec{r}}
\end{align}
where the rotation generated by $H$ is denoted as $Z$ when $\deltat\gg\jhalft$ (rotation axis pointing downwards),
and $R_{\lambda}^{\vec{r}}$ otherwise (rotation by an angle $\lambda$ around an arbitrary axis $\vec{r}$).
Since we have defined a reference energy $\bra{11}H(t)\ket{11}=0$
by canceling the single-qubit dynamic phases (see main text),
and since full population recovery implies that both of the $R$-type rotations pass through the $\ket{11}$ state,
$\langle H \rangle = 0$ is a constant of motion during the two $R$-type rotations.
As a result, the entire contribution from the dynamic phase $\cphase_{\mathrm{D}}$
is contained in the $Z_{\cphase_{\mathrm{D}}}$ rotation.

As a special case, the circular trajectories with $\kphase=0$
(as in \cref{fig:1}e and \cref{fig:1}g, reproduced in \cref{fig:supp_geometrical_phase}a)
only consist of two $R$-type rotations, so that $\cphase_{\mathrm{D}}$ vanishes.
We thus compute the conditional phase from the solid angle $S$ enclosed by a circle on the Bloch sphere:
\begin{align}
	\cphase
	&= \cphase_{\mathrm{G}}
	= -\frac{1}{2}S
	= -\frac{1}{2}2\pi\left(1-\langle\vec{r},\vec{z}\rangle \right)\nonumber\\
	&= -\pi\left(1+\frac{\Delta}{\sqrt{\jhalft^2+\Delta^2}}\right).
\end{align}
This is equivalent to the conditional phase given in \cite{Lacroix2020}
for a gate implemented with a rectangular flux pulse.

As a second special case, we consider the continuous gate set (dashed line in \cref{fig:1}f and trajectory in \cref{fig:1}h, reproduced in \cref{fig:supp_geometrical_phase}b).
In this case, each $R$-type rotation with $\Delta=0$ performs a full population exchange,
yielding $\cphase_{\mathrm{G}}=\pi$.
The $Z$-type rotation happens while the state is $\ket{20}$,
yielding no contribution to the trajectory on the Bloch sphere (no contribution to $\cphase_{\mathrm{G}}$)
and giving a dynamic phase that is equal to the negative time integral of the detuning,
see \cref{eq:geom_phase_nonadiab} combined with \cref{eq:Hsupp} for a $\ket{20}$ state.
This yields the formula used in the main text:
\begin{align}
	\cphase
	=\cphase_{\mathrm{G}}+\cphase_{\mathrm{D}}
	=\pi-\kphase
\end{align}
where we have defined $\kphase$ as the (positive) time integral of the detuning.

We now consider the special case corresponding to the $\cz_{0}$ gates at the top and bottom of \cref{fig:1}f.
An example trajectory with $\cphase_{\mathrm{D}}=\pi$ and $\Delta=0$ is shown in \cref{fig:supp_geometrical_phase}c.
The geometrical phase is
\begin{align}
	\cphase_{\mathrm{G}}
	&=\int_0^{-\pi}\int_0^\lambda \left(-\frac{1}{2}\right) \diff S\nonumber\\
	&=\pi \sin^2{\frac{\lambda}{2}}.
\end{align}
During the $Z_{-\pi}$ rotation (in purple) of duration $t_{Z}=\frac{\pi}{\Delta}$,
we obtain
\begin{align}
	\cphase_{\mathrm{D}}
	&=-\int_0^{t_{Z}} \langle H \rangle \diff t\nonumber\\
	&=-\int_0^\frac{\pi}{\Delta} \Delta \sin^2{\frac{\lambda}{2}} \diff t\nonumber\\
	&=-\pi \sin^2{\frac{\lambda}{2}}\nonumber\\
	&=-\cphase_{\mathrm{G}}.
\end{align}
This yields $\cphase=0$, resulting in a $\cz_{0}$ gate independently of the value of the rotation angle $\lambda$.

We can find an alternative description where the conditional phase is directly reflected by the geometrical phase.
Commuting the $Z$ rotation through the second $R$ gives
\begin{align}
	U = R_{\lambda}^{\vec{r}}Z_{\cphase_{\mathrm{D}}}R_{\lambda}^{\vec{r}}
	=Z_{\cphase_{\mathrm{D}}}R_{\lambda}^{\vec{r}\,'}R_{\lambda}^{\vec{r}}\label{eq:geom_phase_zrr}
\end{align}
where the new rotation axis $\vec{r}\,'$ differs from $\vec{r}$ by the phase $\cphase_{\mathrm{D}}$.
If the time evolution ends in the $\ket{11}$ state,
which is not affected by the pure phase shift on $\ket{20}$,
\cref{eq:geom_phase_zrr} can be simplified as
\begin{align}
	U \equiv R_{\lambda}^{\vec{r}\,'}R_{\lambda}^{\vec{r}}\label{eq:geom_phase_rr}.
\end{align}
The resulting equivalent operation is only composed of $R$-type rotations,
which have a vanishing dynamic phase as explained above.
For the special case considered here,
the trajectory under this equivalent operation is shown in
\cref{fig:supp_geometrical_phase}d,
where $\vec{r}\,'=-\vec{r}$ yields a vanishing solid angle.
The echoing effect of the $Z_{-\pi}$ rotation therefore
produces a $\cz_{0}$ gate with full population recovery and vanishing conditional phase,
for any angle $\lambda$ of the $R$ rotation.

Using the transformation in \cref{eq:geom_phase_rr},
the conditional phase can be interpreted as a geometrical phase
for any gate described by \cref{eq:geom_parameterization},
including the more general cases
visible in \cref{fig:supp_geometrical_space_chevrons} and \cref{fig:supp_geometrical_space_sensitivity_2D}.
\cref{fig:supp_geometrical_phase}e-f provides an example of a generic trajectory,
where the solid angle shown in yellow reflects the conditional phase after the gate.

\section{Sensitivity of the population recovery} \label{supp:param_space_sensitivity}

In the neighborhood of the optimal parameters $\deltamin$ and $\pulselength$,
the sensitivity of the population recovery to these parameters depends on the conditional phase $\cphase$.
We study this effect analytically for small deviations around the optimal parameters.

In the 2-dimensional subspace spanned by $\ket{11}$ and $\ket{20}$,
assuming piecewise-constant net-zero waveforms,
the ideal time evolution is given by three successive unitaries,
as already studied graphically in \cref{supp:geom_phase}:
a partial Rabi rotation $R$ during the time $\pulselength/2$,
a phase rotation $Z_{-\kphase}$, and a second, identical $R$ rotation.

We now derive the corresponding unitaries explicitly,
by rewriting the Hamiltonian \cref{eq:Hsupp} as
\begin{align}
	H(t)
	&= \begin{pmatrix}
		0&\jhalft/2\\
		\jhalft/2&\deltat\\
	\end{pmatrix}\nonumber\\
	&=  \frac{\deltat}{2}I + \frac{\Omega(t)}{2} \left(-\frac{\deltat}{\Omega(t)} \sigma_z + \frac{\jhalft}{\Omega(t)} \sigma_x\right)
\end{align}
where $\Omega(t) = \sqrt{\jhalft^2 + \Delta^2(t)}$ is the effective rotation frequency under $H(t)$.
Fixing $\deltat=\deltamin$ and $\Omega = \sqrt{\jhalft^2 + \deltamin^2}$ yields the $R$ rotation by exponentiating the Hamiltonian using Eq.~(4.8) from \cite{Nielsen2010}:
\begin{align}
	R =& \e{-i H \frac{\pulselength}{2}}\\
	=& \e{-i\frac{\deltamin}{2} \frac{\pulselength}{2}}
	\Bigg(\cos\left(\frac{\Omega}{2}\frac{\pulselength}{2}\right)I\nonumber\\
	&- i\sin\left(\frac{\Omega}{2}\frac{\pulselength}{2}\right) \left(-\frac{\deltamin}{\Omega}\sigma_z +\frac{\jhalft}{\Omega}\sigma_x \right) \Bigg).
\end{align}
Fixing instead $\deltat\gg\jhalft$ yields the phase rotation:
\begin{align}
	&Z_{-\kphase} = \ket{11}\!\bra{11} + \e{-i\kphase}\ket{20}\!\bra{20}.
\end{align}

\begin{figure}
	\centering
	\includegraphics[width=\linewidth]{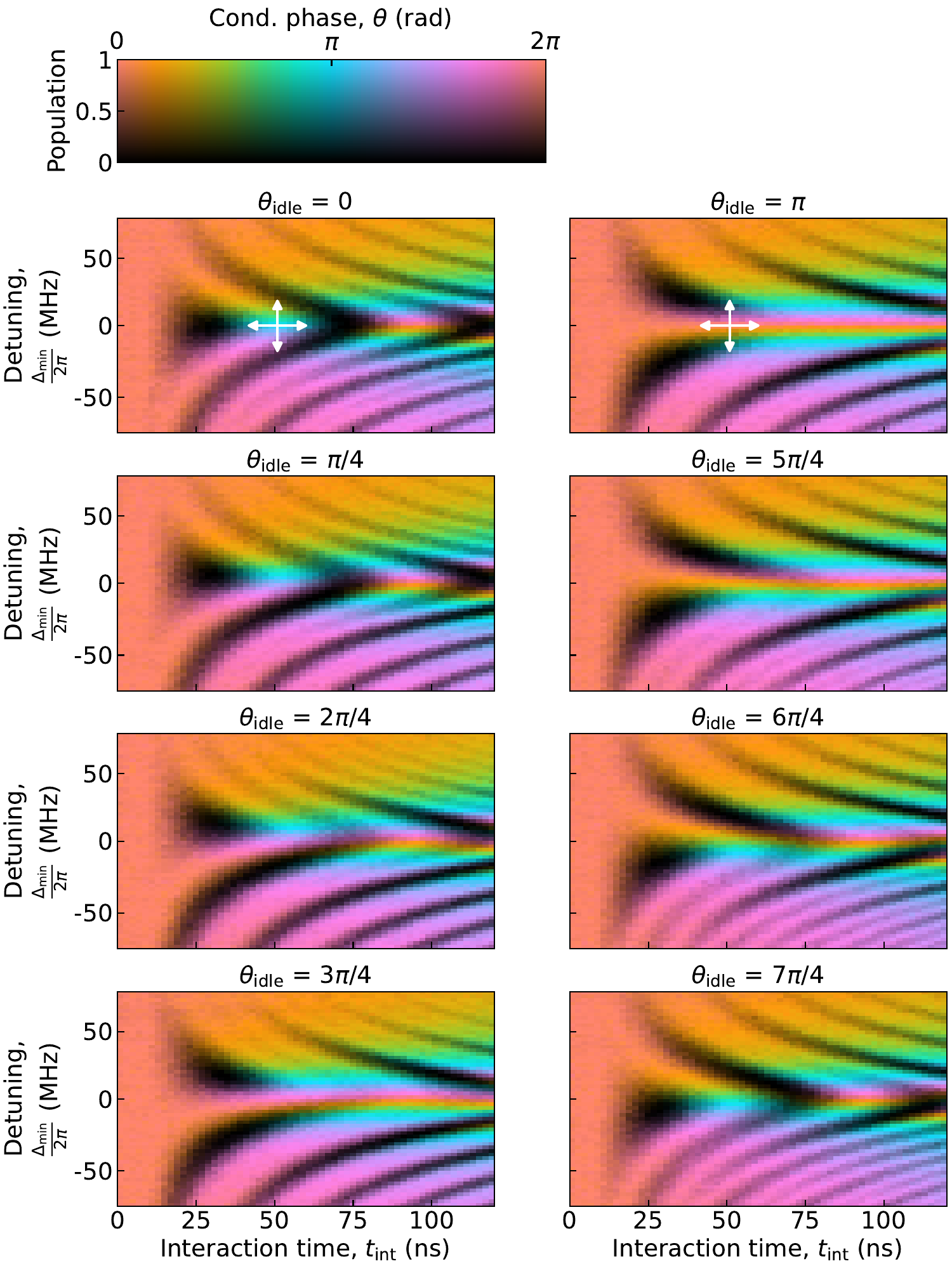}
	\caption{
		\figtitle{Measured $\ket{11}$ population and conditional phase after one gate},
		for different values of the idle phase $\kphase$.
		The white arrows indicate the one-dimensional cuts shown in \cref{fig:supp_geometrical_space_sensitivity_1D}
		and approximated in \cref{eq:taylor}.
	}
	\label{fig:supp_geometrical_space_chevrons}
\end{figure}
The final probability amplitude of the $\ket{11}$ state is then the matrix element:
\begin{align}
	\label{eq:amp11}
	c_{\ket{11}} = &\bra{11}RZ_{-\kphase}R\ket{11} =
	\e{-i\frac{\deltamin}{2}\pulselength}\times\\
	&\left(\xi +  \left(1 - \xi \right)\cos\left(\frac{\Omega}{2}\pulselength\right) + i \frac{\deltamin}{\Omega} \sin\left(\frac{\Omega}{2}\pulselength\right)  \right)\nonumber
\end{align}
with $\xi = (1 - \e{-i\kphase}) \jhalft^2 / 2\Omega^2$ accounting for the phase rotation in the middle of the evolution.
This matrix element is a function of three controllable parameters: $\deltamin$, $\pulselength$ and $\kphase$.
In \cref{fig:supp_geometrical_space_chevrons}, we show measurement data for
the population recovery $P_{\ket{11}}=|c_{\ket{11}}|^2$ (encoded as brightness) and the conditional phase $\arg c_{\ket{11}}$ (color),
as a function of $\deltamin$ and $\pulselength$, for eight values of $\kphase$.
The cases $\kphase=0$ and $\kphase=\pi$ correspond to \cref{fig:1}e and \cref{fig:resonance}a, respectively.
The other panels illustrate the continuous transition between these two extreme cases.

\begin{figure}
	\centering
	\includegraphics[width=\linewidth]{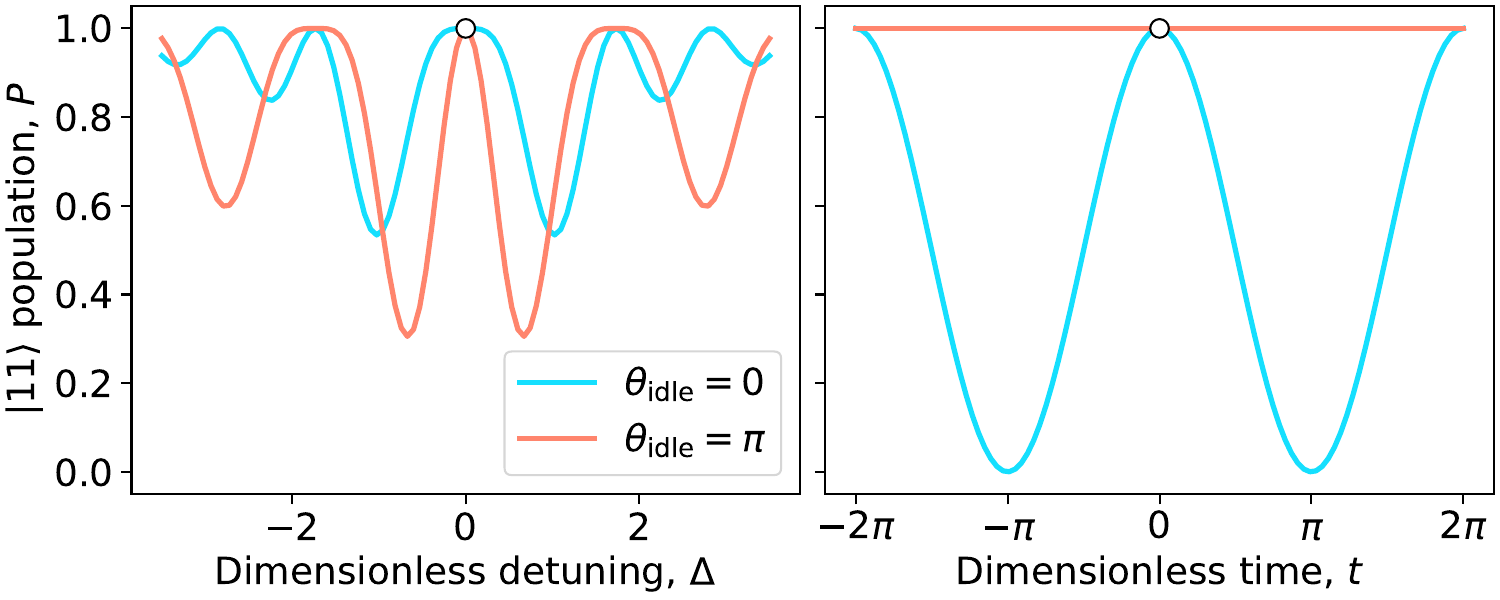}
	\caption{
		\figtitle{Simulated $\ket{11}$ population after one gate},
		using \cref{eq:amp11}.
		The population is shown
		as a function of the dimensionless parameters
		$\Delta=\deltamin/\jhalft$ and $t=\pulselength\jhalft-2\pi$,
		around the $\cz_{\pi}$ gate ($\kphase=0$) and the $\cz_{0}$ gate ($\kphase=\pi$).
		The continuous gate set, corresponding to both parameters set to 0, is indicated as a white dot.}
	\label{fig:supp_geometrical_space_sensitivity_1D}
\end{figure}

We now study the behavior of the population recovery in the neighborhood of the continuous gate set.
Below, we use dimensionless parameters $\Delta=\deltamin/\jhalft$ and $t=\pulselength\jhalft-2\pi$.
Note the $2\pi$ offset in $t$,
such that the ideal continuous gate set corresponds to unitaries with both $\Delta=0$ and $t=0$.
\cref{fig:supp_geometrical_space_sensitivity_1D} shows the population recovery $P_{\ket{11}}=|c_{\ket{11}}|^2$,
with $c_{\ket{11}}$ calculated from \cref{eq:amp11},
as a function of either parameter, both at $\kphase=0$ and $\kphase=\pi$.

Taylor expanding $P_{\ket{11}}(\Delta,t)$ around 0 yields, after a few lines of algebra,
\begin{align}
	P_{\ket{11}}(\Delta, 0) &=
	\begin{cases}
		1 - \frac{\pi^{2} \Delta^{4}}{4} + O\left(\Delta^{6}\right), & \text{if}\ \kphase=0\\
		1 - 4 \Delta^{2} + O\left(\Delta^{4}\right), & \text{if}\ \kphase=\pi
	\end{cases}\nonumber\\
	P_{\ket{11}}(0, t) &=
	\begin{cases}
		1 - \frac{t^{2}}{4} + O\left(t^{4}\right), & \text{if}\ \kphase=0\\
		1, & \text{if}\ \kphase=\pi.
	\end{cases}
	\label{eq:taylor}
\end{align}
These four cases are indicated by the white arrows in \cref{fig:supp_geometrical_space_chevrons}.
The population recovery is only fourth-order sensitive to the detuning $\Delta$ at $\kphase=0$.
Intuitively, around the $\cz_{\pi}$ gate (see \cref{fig:supp_geometrical_space_chevrons} at $\kphase=0$),
small changes of $\Delta$ are tangent to the curve of maximal population recovery,
and therefore have very little effect on the population.
On the other hand, the population recovery is second-order sensitive to $\Delta$ at $\kphase=\pi$.
This is why we fine-tune the detuning at $\kphase=\pi$, as discussed in the main text.
Similarly, we fine-tune the interaction time $\pulselength$ at $\kphase=0$,
noting that $\kphase=\pi$ would yield no sensitivity to $\pulselength$ at all.
This insensitivity can be seen from \cref{eq:taylor} and can be interpreted as the echo effect discussed in \cref{supp:geom_phase}.

\begin{figure}
	\centering
	\includegraphics[width=\linewidth]{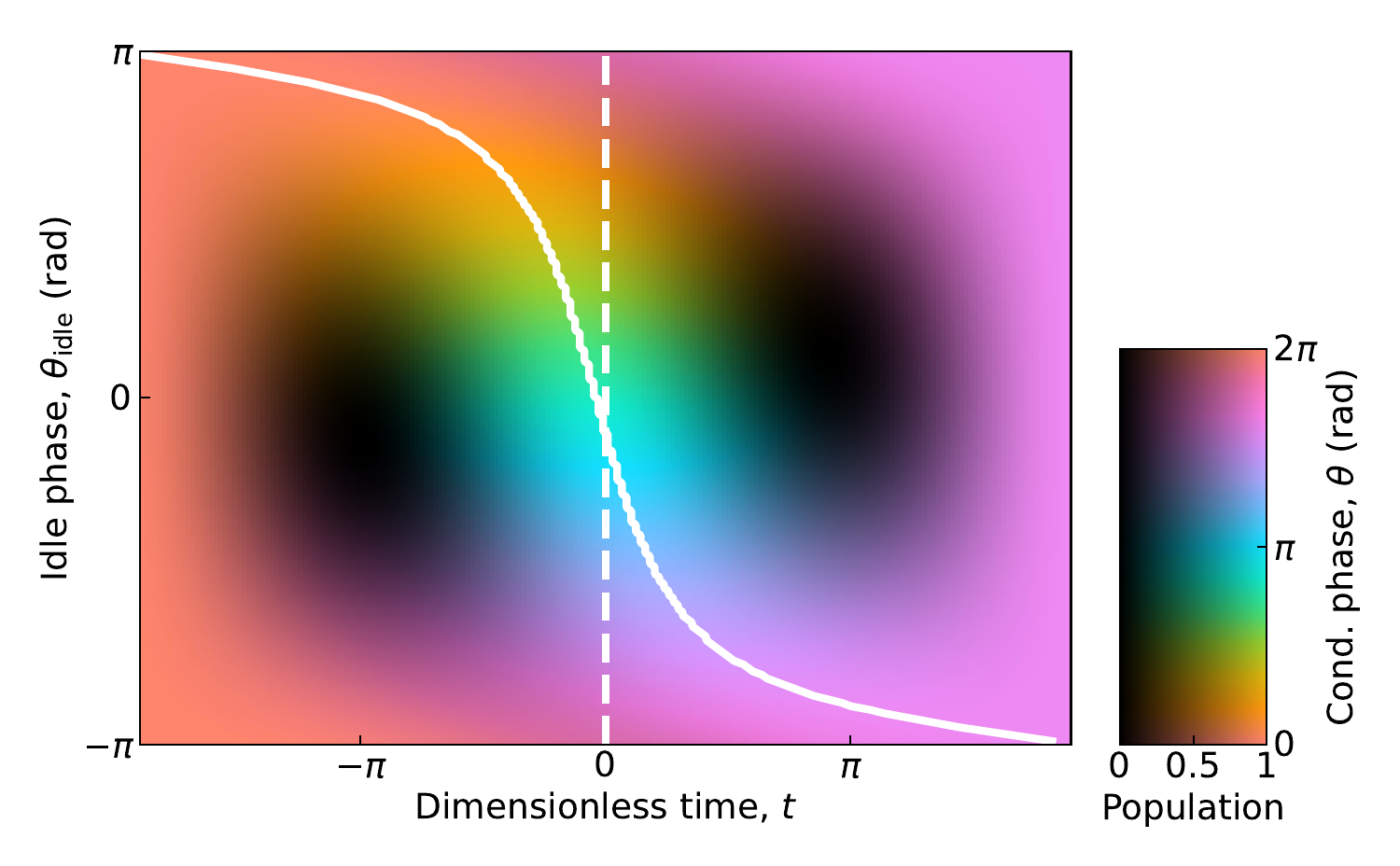}
	\caption{
		\figtitle{Simulated $\ket{11}$ population and conditional phase after one gate},
		for an exaggeratedly high value of the dimensionless detuning, $\Delta=0.2$,
		as a function of the idle phase $\kphase$ and the dimensionless time $t$.
		The curve of maximal population recovery is shown as a solid white line.
		The dashed white line indicates the curve of maximal population recovery if $\Delta$ was instead equal to $0$.
	}
	\label{fig:supp_geometrical_space_sensitivity_2D}
\end{figure}

We now quantitatively study the effect of an imperfect calibration
of the detuning $\Delta$,
since fine-tuning the detuning is a requirement to reach the region of the parameter space shown in \cref{fig:1}f,
where the effects of the interaction time and idle time are decoupled.
\cref{fig:supp_geometrical_space_sensitivity_2D} shows an example of population recovery and conditional phase at a nonzero detuning, yielding a skewed landscape compared to \cref{fig:1}f.
The unrealistically high value of the detuning, $\Delta=0.2$,
is chosen for the purpose of illustration.
The solid white curve indicates the points of maximal population recovery, evaluated numerically.
This would correspond to a continuous gate set parameterized by both the idle phase and the interaction time.
For decreasing $|\Delta|$, this curve approaches the vertical dashed line at $t=0$, parameterized by the idle phase only.
For a nonzero $|\Delta|$, a continuous gate set parameterized only by the idle phase therefore cannot reach zero leakage for all gates.

We now study such a continuous gate set parameterized by $\kphase$ only.
We assume for concreteness $t=0$, such that the population recovery will be lowest at $\kphase=\pm\pi$ (the extremities of the dashed white line in \cref{fig:supp_geometrical_space_sensitivity_2D}).
At this point, the leakage is $1-P_{\ket{11}}(\Delta, 0) \approx 4 \Delta^2$
due to \cref{eq:taylor}.
The spacing of $\SI{33}{\micro\Phi_0}$ between points used in the calibration in \cref{fig:resonance}d
corresponds to a detuning of $4\times 10^{-3}$
and would, therefore, allow leakage below $4 \Delta^2\approx 7\times 10^{-5}$
in the absence of all other imperfections.

\section{Leakage amplification} \label{supp:la}

In the calibration procedure of the continuous gate set,
we use the fact that
the leakage after a series of $N$ gates displays interference fringes as a function of the gate separation.
We first derive this behavior analytically in a simplified model by assuming a unitary process,
and then quantitatively in simulations by fitting a master equation model to the measured data.

Coherent amplification of the residual state population due to off-resonant Rabi oscillations was studied analytically in \cite{Vitanov2020}.
To obtain constructive interferences,
this work suggested concatenating pairs of gates which have an opposite effect on the
phase of the final state.
This method was, for example,
implemented in a procedure for benchmarking leakage of single-qubit gates in \cite{Chen2016},
although the resulting measurement does not distinguish between the leakage of each gate within the pair.
The ideas from \cite{Vitanov2020} were extended to a driven qutrit system in \cite{Xu2023b},
and used as part of the single-qubit gate calibration in an experiment in \cite{Underwood2024}.
Error amplification has been more generally applied in order to estimate coherent gate errors
in superconducting qubits, by repeating single-qubit \cite{Sheldon2015, Asaad2016, Lazar2023} or two-qubit gates \cite{Marxer2022}.

In cases where the errors to be characterized add up incoherently,
for example in randomized benchmarking \cite{Magesan2012a, Magesan2012},
the resulting amplified error scales linearly with the number of gates.
If, on the other hand, the errors add coherently,
the resulting interferences must be constructive in order to observe an amplified error.
In the case of single-qubit gates, this can be done by adjusting the phase of each gate to follow the rotating frame of the qubit \cite{McKay2017}.
For the two-qubit gate implementation in this paper,
this degree of freedom allowing to adjust the phase of the gate does not exist.
We thus, instead,
adjust the gate separation in a sequence of two-qubit gates to cancel the dynamic phase accumulated by the leaked population,
resulting in constructively interfering total leakage.

\begin{figure}
	\centering
	\includegraphics[width=\linewidth]{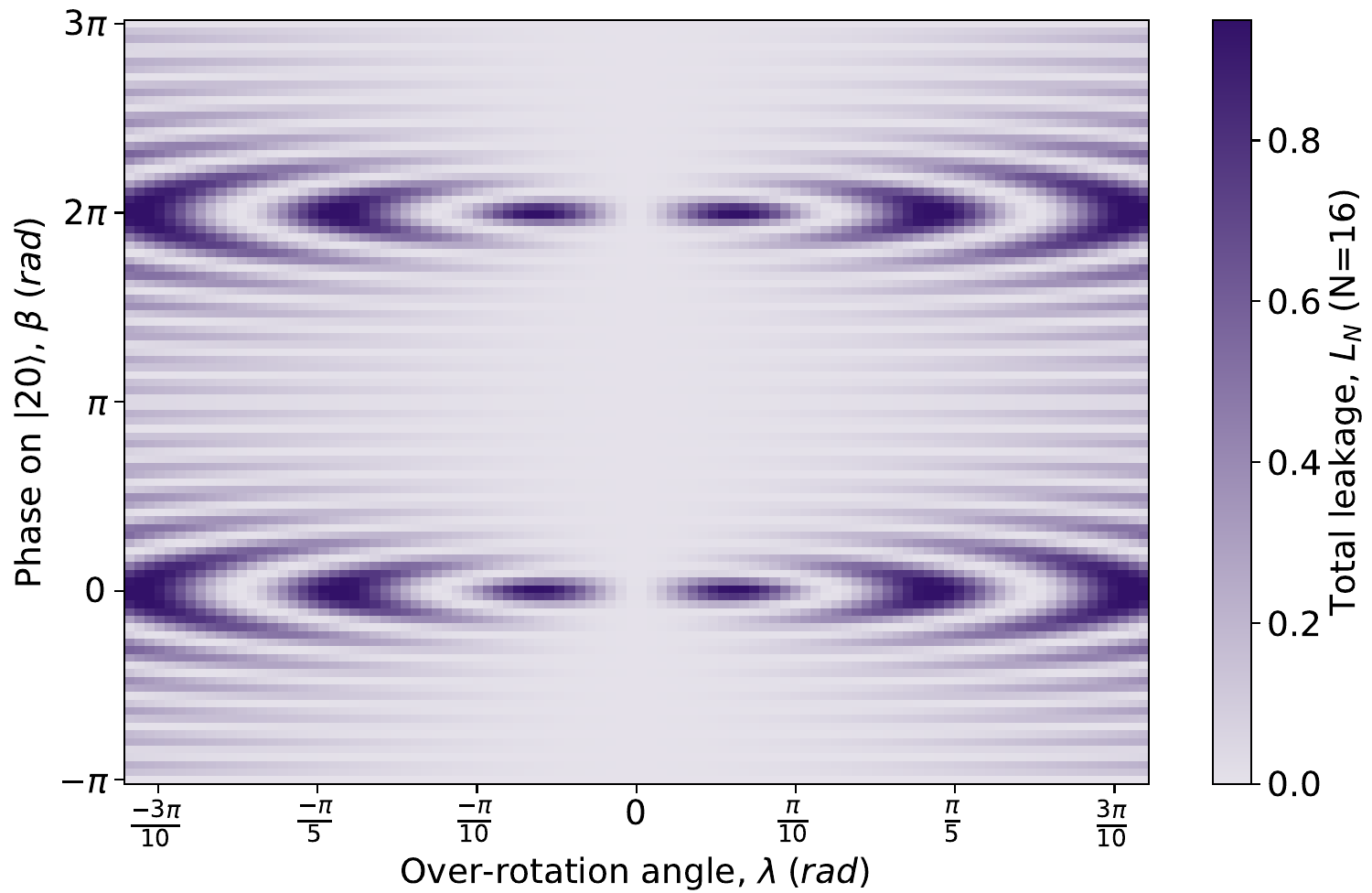}
	\caption{
		\figtitle{Calculated interference pattern}
		of the coherent leakage after $N=16$ gates
		using the simplified decoherence-free model in \cref{eq:LA_full}.
		The leakage is shown
		as a function of the over-rotation angle $\lambda$, which sets the strength of the leakage,
		and of the phase acquired by the population in the $\ket{20}$ state between two gates,
		which sets the interference condition.
	}
	\label{fig:sim_leakage_amp}
\end{figure}

We start by deriving the leakage after $N$ gates analytically,
by assuming fully coherent leakage.
We restrict the description to the subspace spanned by $\ket{11}$ and $\ket{20}$.
We model an entire gate as the unitary
\begin{align}
	\label{eq:LA_coh_unitary}
	U = Y_{\lambda}Z_{\beta}
	= \begin{pmatrix}
		\cos(\frac{\lambda}{2})&-\sin(\frac{\lambda}{2})\\
		\sin(\frac{\lambda}{2})&\cos(\frac{\lambda}{2})
	\end{pmatrix}
	\begin{pmatrix}
		\e{-i\frac{\beta}{2}}&\\
		&\e{i\frac{\beta}{2}}
	\end{pmatrix}
\end{align}
parameterized by $\lambda$ and $\beta$.
Compared to a fully generic unitary $\e{i\alpha}Z_{\beta_1}Y_{\lambda}Z_{\beta_2}$ (Theorem 4.1 in \cite{Nielsen2010}),
we have dropped the global phase
and have absorbed a $Z$ rotation into the (arbitrary) choice of the phase reference for the $\ket{20}$ state.
In the parameterization in \cref {eq:LA_coh_unitary}, $\lambda$ encompasses the contributions of any coherent rotation errors
and $\beta$ represents the phase acquired due to the separation $\bufferlength$ between successive gates.
Ideally, we have $\lambda=0$ for a calibrated gate
and $\beta=0$ to obtain constructive interferences.

Expressing $U^N$ using Eq.~(5) from \cite{Vitanov2020} yields
\begin{align}
	\bra{20}U^N\ket{11} &= \bra{20}U\ket{11} \frac{\sin\left(N\arccos{\mathfrak{Re}[\bra{11}U\ket{11}]}\right)}{\sin\left(\arccos{\mathfrak{Re}[\bra{11}U\ket{11}]}\right)}
\end{align}
and taking the square modulus gives the total leakage after $N$ gates
\begin{align}
	\leakage_N
	&= \underbrace{\sin^2\left(\frac{\lambda}{2}\right)}_{\leakage_1} \frac{\sin^2\left(N\arccos\left({\cos\left(\frac{\lambda}{2}\right)\cos\left(\frac{\beta}{2}\right)}\right)\right)}{\sin^2\left(\arccos\left({\cos\left(\frac{\lambda}{2}\right)\cos\left(\frac{\beta}{2}\right)}\right)\right)}\label{eq:LA_full}.
\end{align}
The interference pattern described by \cref{eq:LA_full}
is plotted in \cref{fig:sim_leakage_amp} as a function of $\lambda$ and $\beta$.

\cref{eq:LA_full} can be rewritten to make the dependence on the leakage of a single gate $\leakage_1$ explicit:
\begin{align}
	\label{eq:LA_full_L1}
	\leakage_N &=
	\leakage_1 \underbrace{\frac{\sin^2\left(N\arccos\left({\cos\left(\arcsin\left(\sqrt{\leakage_1}\right)\right)\cos\left(\frac{\beta}{2}\right)}\right)\right)}{\sin^2\left(\arccos\left({\cos\left(\arcsin\left(\sqrt{\leakage_1}\right)\right)\cos\left(\frac{\beta}{2}\right)}\right)\right)}}_{f(\leakage_1, \beta)}.
\end{align}

\begin{figure}
	\centering
	\includegraphics[width=\linewidth]{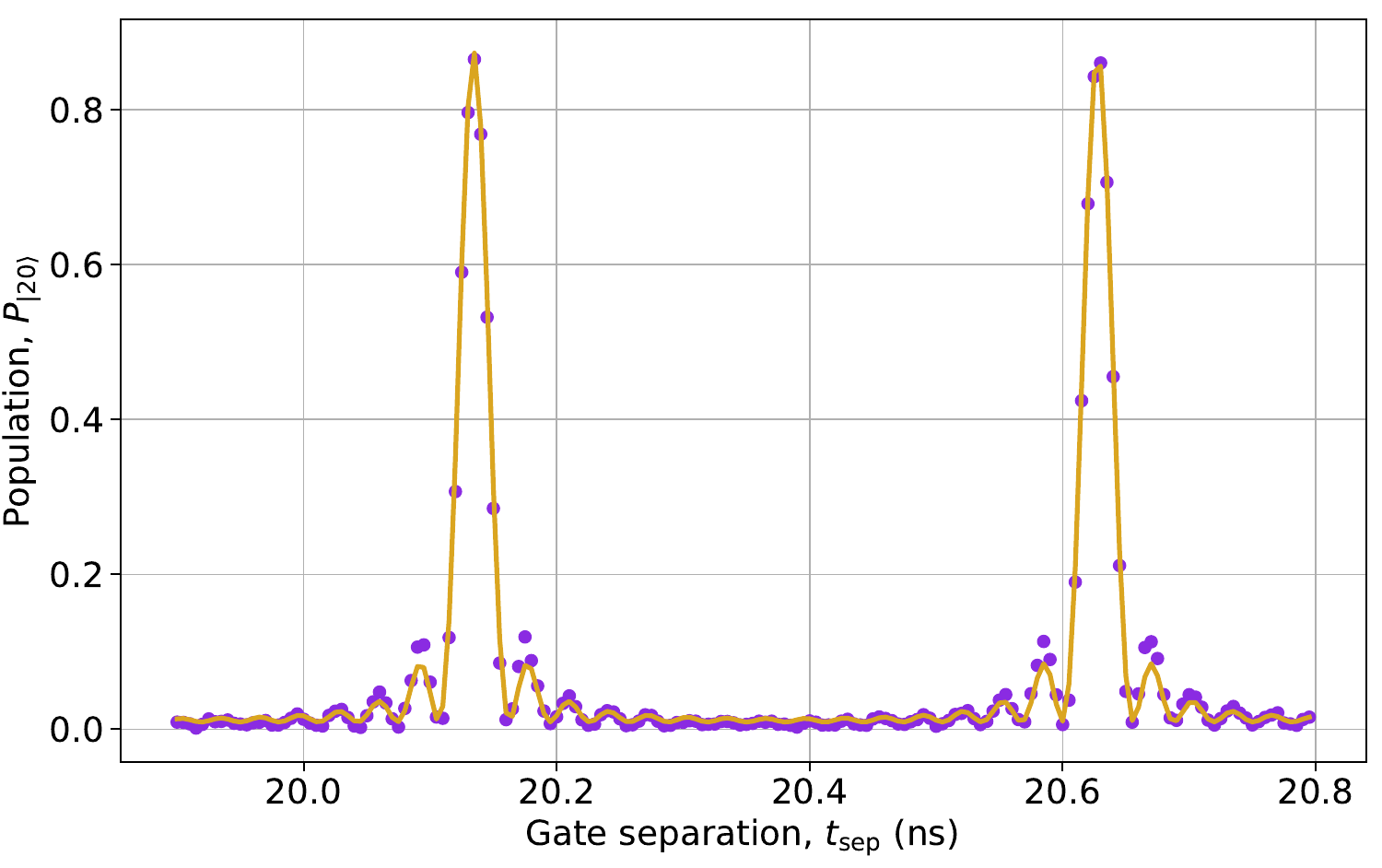}
	\caption{
		\figtitle{Amplified leakage}
		(dots)
		along a one-dimensional cut in the dataset shown in \cref{fig:resonance}c,
		at the point of highest contrast near $\fluxl=\SI{-0.1996}{\Phi_0}$,
		and fit (line) using the model in \cref{eq:LA_full_model}.
	}
	\label{fig:supp_fit_leakage_amplification}
\end{figure}

In the special case of small leakage $\leakage_1$ per gate,
see the range around $\lambda=0$ in \cref{fig:sim_leakage_amp}
and the experimental data in \cref{fig:resonance}c,
we can approximate \cref{eq:LA_full_L1} as its first-order Taylor expansion in $L_1$:
\begin{align}
	\leakage_N &= \leakage_1 f(0, \beta) + O\left(\leakage_1^2\right)\nonumber\\
	&\approx \leakage_1\frac{\sin^2(\frac{N\beta}{2})}{\sin^2(\frac{\beta}{2})}
	= \leakage_1\frac{1-\cos(N\beta)}{1-\cos(\beta)}.
\end{align}
The resulting expression can be interpreted as
the interference pattern of $N$ waves of equal amplitude $\leakage_1$,
each shifted by an integer multiple of the phase $\beta$ accumulated after each gate.
This explains the interference fringes visible as a function of $\beta$
along a vertical cut in \cref{fig:resonance}c or \cref{fig:sim_leakage_amp},
see an example in \cref{fig:supp_fit_leakage_amplification}.

In the special case of constructive interference ($\beta=0$),
the denominator in \cref{eq:LA_full_L1} cancels with the prefactor $\leakage_1$.
For small leakage $\leakage_1$,
plugging in the Taylor expansion
$\leakage_1 = 1-P_{\ket{11}}  \approx 4 \Delta^2 = 4 \deltamin^2/\jhalft^2$
from \cref{supp:param_space_sensitivity} yields
\begin{align}
	\leakage_N &\approx \sin^2\left(2N\frac{\deltamin}{\jhalft}\right).
\end{align}
With a first-order approximation of $\deltamin$ as a function of the pulse parameter $\fluxl$,
this explains the sinusoidal shape of the data shown in \cref{fig:resonance}d as a function of $\fluxl$.

We now study the total leakage quantitatively,
by computing
\begin{align}
	\label{eq:LA_full_model}
	\leakage_N = \mathrm{Tr}\left[\ket{20}\!\bra{20}U^N(\ket{11}\!\bra{11})\right]
\end{align}
numerically in a master equation simulation \cite{AmShallem2015}.
We model the time evolution
in an effective three-dimensional Hilbert space including
the $\ket{11}$ and $\ket{20}$ states and an auxiliary state representing all lower-energy states.
We take decoherence into account by including Lindblad terms describing
decay from $\ket{11}$ to the lower-energy states with rate $T_1^{\ket{11}}$,
decay from $\ket{20}$ to the lower-energy states with rate $T_1^{\ket{20}}$,
and dephasing between $\ket{11}$ and $\ket{20}$ with rate $T_{\phi}$.
We then truncate the Hilbert space and the operators to the $\ket{11}$--$\ket{20}$ subspace.

We model the time evolution assuming piecewise-constant, net-zero waveforms.
To approximate the increased sensitivity to flux noise during the frequency excursions \cite{Krinner2022}
while keeping a limited number of parameters,
we only apply the dephasing term during the interaction time,
whereas we apply the relaxation terms during the entire time evolution.
From the experimentally determined relaxation times
in \cref{tab:device_params} we obtain
$T_1^{\ket{11}} = ( 1/T_1^{\ket{10}} + 1/T_1^{\ket{01}} )^{-1} =\SI{23.8}{\micro\second}$
and
$T_1^{\ket{20}}=\SI{14.7}{\micro\second}$.
This is an approximation, which neglects the frequency dependence of the relaxation times.
Plugging in measured values for the dephasing time
would require to account for the strong frequency dependence of
the sensitivity to flux noise
and for the echo effect coming from the net-zero pulse \cite{Rol2019}.
We thus take the simpler approach of only considering an effective dephasing time $T_{\phi}$,
which we fit to the data.
As further fit parameters,
we choose the detuning $\deltamin$ during the interaction time
(responsible for coherent errors),
and a time offset $t_{\mathrm{offset}}$ to account for the rise times of the control pulses
by expressing $\beta$ in \eqref{eq:LA_coh_unitary} as $\beta = (\omega_{20}-\omega_{11})(\bufferlength-t_{\mathrm{offset}})$
with the gate separation $\bufferlength$.

In order to extract the leakage of a single gate $\leakage_1$ from the total leakage $\leakage_N$,
we fit the model in \cref{eq:LA_full_model} to a one-dimensional subset of the dataset shown in \cref{fig:resonance}c-d,
as a function of the gate separation $\bufferlength$,
for a fixed value of the flux amplitude near $\fluxl=\SI{-0.1996}{\Phi_0}$ yielding the highest contrast.
The fit as a function of the gate separation
is shown in \cref{fig:supp_fit_leakage_amplification}.
We obtain the fitted value $T_{\phi} = \SI{18\pm2}{\micro\second}$.
For comparison, from a model approximating the noise power spectral density as white,
using Eqs. (19) and (A3) from \cite{Strauch2025},
this effective dephasing time would correspond to an incoherent leakage
$\leakage_1 \approx \pulselength/4 T_{\phi} = 6.9\times 10^{-4}$.
This is of the same order of magnitude as the leakage measured in \cref{fig:benchmarking}d
for a calibrated gate.

Keeping $T_{\phi}$ and $t_{\mathrm{offset}}$ fixed,
we now reuse the fitted model for the rest of the dataset, as follows.
We compute both $\leakage_N$ and $\leakage_1$ as a function of $\deltamin$,
and interpolate numerically $\leakage_N = f(\leakage_1)$.
We use $f$ to compute the right-hand vertical axis in \cref{fig:resonance}d.
The area which is not shaded in red indicates the range
where this conversion can be done in a unique way.

\section{Cross-entropy benchmarking} \label{supp:xeb}

We use cross-entropy benchmarking (XEB), introduced in \cite{Boixo2018, Arute2019},
to estimate the fidelity of a gate or the whole continuous set of gates.
XEB does not require a set of gates forming a group,
as needed in interleaved randomized benchmarking \cite{Magesan2012a, Magesan2012},
thus providing more flexibility in the choice of gates to be characterized.
Unlike state tomography \cite{Steffen2006a},
XEB evaluates random circuits containing many gates
and of different lengths, thereby canceling the contribution
of state preparation and measurement errors,
and allowing to capture errors that only arise in long gate sequences
such as non-Markovian effects due to long-time control pulse distortions.
XEB has been applied to benchmark entangling two-qubit gates, for example in \cite{Arute2019}.
Alternatively, a continuous gate set was characterized using randomized benchmarking in \cite{Hill2021, Abrams2020},
by benchmarking the Clifford unitary $\cz_{\cphase}\cz_{\pi-\cphase}$,
although this method does not distinguish the individual errors of each gate of the pair of gates.

An XEB experiment consists in measuring the statistics of observing each $n$-qubit basis state,
for random circuits with a varying number $M$ of gate cycles.
An unbiased estimate of the fidelity per cycle can be computed
under the assumptions of low error rates and chaotic state evolution \cite{Boixo2018, Arute2019, Helsen2022a, Proctor2017}.
In this context, chaotic evolution means that a few cycles are sufficient to form a 2-design,
that is, the distribution over random circuits resembles the uniform distribution
up to second-order statistical moments.
This is the reason for adding the mixing unitary $\umix$ in the cycle design shown in \cref{fig:benchmarking}b.
From the output of the randomized circuits,
one then estimates the XEB depolarization fidelity \cite{Boixo2018, Foxen2020},
which is a normalized function of the cross-entropy between the ideal (simulated) and measured probability distributions.
The XEB fidelity decreases exponentially with the number of cycles $M$ \cite{Arute2019, Lazar2023a}:
\begin{align}
	\label{eq:fxeb}
	F_{\mathrm{XEB}} = A\left(1-\frac{D}{D-1}\epsilon_a\right)^M+B
\end{align}
where $\epsilon_a$ is the average error per cycle, $D$ is the size of the Hilbert space ($D=4$ in our case), and $A, B$ represent contributions of state preparation and measurement errors.
The average cycle error $\epsilon_a$ can be extracted from an exponential fit of $F_{\mathrm{XEB}}$ as a function of $M$.

In an XEB experiment,
we run 250 different random circuits consisting of $\umix$ and the gate under test (\cref{fig:benchmarking}b)
for each number of cycles $M\in\{4, 8, 16, 32, 64, 128, 256\}$, each repeated 2048 times.
We perform three-level single-shot readout
to detect leakage in the final readout, see \cref{fig:benchmarking}d.
After excluding the $\SI{3.6}{\%}$ of the experimental runs where leakage was detected in the final readout,
we renormalize the remaining data
such that the populations sum to one in the computational subspace, as in \cite{Lazar2023}.
We fit the exponential decay of the average XEB depolarization fidelity and extract the average cycle error.
In addition, we perform a reference experiment benchmarking only $\umix$.
We finally extract the error $\epsilon_{\cz}$ of a gate under test
by writing, similarly to the product over cycles in \cref{eq:fxeb},
the product of fidelities within a cycle
$1-\epsilon_{\mathrm{tot}} \approx (1-\epsilon_{\cz})(1-\epsilon_{\umix})$,
where $\epsilon_{\mathrm{tot}}$ and $\epsilon_{\umix}$ are the average cycle errors
for the measurements with and without the gate under test, respectively.

%\bibliography{}

\bibliography{F:/Writing/RefDB/QudevRefDB.bib}

\end{document}